\documentclass[twoside,11pt,a4paper,parskip=half]{scrartcl}
\usepackage[utf8]{inputenc}
\usepackage[T1]{fontenc}
\usepackage{textgreek}
\usepackage{textcomp}
\usepackage{amsmath,amssymb}
\usepackage{a4wide}
\usepackage{soul}
\usepackage[dvipsnames]{xcolor}
\usepackage{graphicx}
\usepackage{multirow}
\usepackage{hyperref}
\usepackage{tabularx}
\usepackage{mathtools}
\usepackage{xspace}
\usepackage[separate-uncertainty,multi-part-units=single,list-units=single]{siunitx}
\usepackage{collcell,booktabs}
\usepackage[capitalize]{cleveref}
\usepackage[citestyle=numeric-comp,bibstyle=HBnumeric,sorting=none]{biblatex}
\usepackage{multicol}

\renewcommand{\arraystretch}{1.2}

\newcommand{\ie}{i.e.~}
\newcommand{\eg}{e.g.~}

\newcommand{\CL}[1]{\ensuremath{\SI{#1}{\percent}~\text{C.L.}}}
\newcommand{\CLs}{\ensuremath{\text{CL}_s}\xspace}
\newcommand{\Code}[1]{\texttt{\detokenize{#1}}}
\newcommand{\NEW}{\textcolor{NavyBlue}{(NEW)}\xspace}

\newcommand{\HB}{\texttt{HiggsBounds}\xspace}
\newcommand{\HS}{\texttt{HiggsSignals}\xspace}
\newcommand{\HSv}[1]{\texttt{HiggsSignals-#1}}
\newcommand{\HBv}[1]{\texttt{HiggsBounds-#1}}

\newcommand{\vhatnnlotwo}{\texttt{VH@NNLO-2.0}}
\newcommand{\vhatnnlo}{\texttt{VH@NNLO}}

\newcommand{\gray}[1]{\textcolor{gray}{#1}}
\newcommand{\sw}{\sin\theta_\text{w}}
\newcommand{\cw}{\cos\theta_\text{w}}
\newcommand{\iu}{{i\mkern2mu}}
\newcommand{\eqcomma}{\,,}
\newcommand{\eqdot}{\,.}

\DeclareSIUnit{\TeV}{TeV}
\DeclareSIUnit{\pb}{pb}
\DeclareSIUnit{\fb}{fb}

\newcolumntype{P}[1]{>{\raggedleft\collectcell\Code}p{#1}<{\endcollectcell}}
\newcolumntype{R}{>{\collectcell\Code}r<{\endcollectcell}}

\graphicspath{{figs/}}

\begin{document}

\thispagestyle{empty}

\mbox{}\hfill\texttt{BONN-TH-2020-03},
\texttt{DESY 20-093},
\texttt{IFT-UAM/CSIC-20-072},
\texttt{LU 20-27}

\def\thefootnote{\fnsymbol{footnote}}

\begin{center}
	\Large\textbf{\boldmath
		\HBv{5}: Testing Higgs Sectors in the LHC 13 TeV Era
		\unboldmath}
\end{center}
\vspace{-0.5cm}
\begin{center}
	Philip Bechtle$^{1}$, Daniel Dercks$^{2}$\footnote{Former affiliation.}, Sven Heinemeyer$^{3,4,5}$, Tobias Klingl$^1$, \\[.5em]
	Tim Stefaniak$^{2}$, Georg Weiglein$^{2}$ and Jonas Wittbrodt$^{6}$\footnote{Electronic addresses:
		bechtle@physik.uni-bonn.de,
		sven.heinemeyer@cern.ch,
		klingl@physik.uni-bonn.de,\\
		tim.stefaniak@desy.de,
		georg.weiglein@desy.de,
		jonas.wittbrodt@thep.lu.se} \\[0.4cm]
	{\small
	\textsl{$^1$Physikalisches Institut der Universit\"at Bonn,
		Nu{\ss}allee 12, D-53115 Bonn, Germany}\\[0.1cm]
	\textsl{ $^2$ Deutsches Elektronen-Synchrotron DESY,
		Notkestra{\ss}e 85, D-22607 Hamburg, Germany}\\[0.1cm]
	\textsl{$^3$Campus of International Excellence UAM+CSIC, Cantoblanco, E--28049 Madrid, Spain}\\[0.1cm]
	\textsl{$^4$Instituto de F\'isica Te\'orica, (UAM/CSIC), Universidad
		Aut\'onoma de Madrid,\\ Cantoblanco, E-28049 Madrid, Spain
	}\\[0.1cm]
	\textsl{$^5$Instituto de F\'isica de Cantabria (CSIC-UC), E-39005 Santander,
		Spain}\\[0.1cm]
	\textsl{$^6$Department of Astronomy and Theoretical Physics, Lund University, SE-22100 Lund, Sweden}\\[1mm]
	}
\end{center}
\vspace{0.2cm}

\renewcommand{\thefootnote}{\arabic{footnote}}
\setcounter{footnote}{0}

\begin{abstract}
	We describe recent developments of the public computer code \HB.
	In particular, these include the incorporation of LHC Higgs search results
	from Run 2 at a center-of-mass energy of \SI{13}{\TeV}, and an updated and extended framework
	for the theoretical input that accounts for improved Higgs cross
	section and branching ratio predictions and new search channels. We furthermore discuss an improved method used in \HB\ to approximately reconstruct the exclusion likelihood for LHC searches for
	non-standard Higgs bosons decaying to $\tau\tau$ final states.  We describe
	in detail the new and updated functionalities of the new version \HBv{5}.
\end{abstract}

\clearpage

\tableofcontents

\clearpage

\section{Introduction}
\label{sec:intro}

After the discovery of a Higgs boson with a mass around
\SI{125}{\GeV}~\cite{Chatrchyan:2012xdj,Aad:2012tfa} the searches for new
scalars have intensified and expanded into more and more search signatures.
Evidence for the existence of additional neutral and/or electrically charged
Higgs bosons would be an unambiguous sign of physics beyond the Standard Model
(BSM), where the SM scalar sector is extended by new scalar field(s). These
fields could be singlets, doublets or even higher representations of the
electroweak gauge group $\text{SU}(2)_L$. Well-known examples of such extensions
are the real or complex scalar singlet
extension~\cite{Schabinger:2005ei,Patt:2006fw,Barger:2007im,Barger:2008jx},
featuring one or two additional neutral scalar bosons, respectively, and the
Two-Higgs-Doublet Model (2HDM)~\cite{Gunion:1989we,Gunion:2002zf,Branco:2011iw},
which contains three neutral Higgs bosons --- typically denoted $h, H$ and $A$
in the CP-conserving case --- and a pair of charged Higgs bosons, $H^\pm$. The
Higgs sector of the Minimal Supersymmetric Standard Model (MSSM), for instance,
is at the tree-level a specific version of a
2HDM~\cite{Gunion:1986nh,Martin:1997ns,Carena:2002es}. Examples of BSM Models
containing both additional scalar doublets and singlets are the
Next-to-2HDM~\cite{Muhlleitner:2016mzt,Chen:2013jvg} and the
Next-to-MSSM~\cite{Ellwanger:2009dp,Maniatis:2009re}, while higher
representations of scalars are considered \eg in the Georgi-Machacek
model~\cite{Georgi:1985nv}.

Since --- so far --- no additional Higgs bosons have been discovered at the LHC
all of these searches have resulted in exclusion limits that constrain the
possible parameter space of BSM theories with extended Higgs
sectors. Due to the large number of results in many different search channels,
the task of testing BSM model predictions against the assembled results from
Higgs searches warrants dedicated tools. The tool \HB~\cite{Bechtle:2008jh,
	Bechtle:2011sb, Bechtle:2013wla, Bechtle:2015pma} has been developed to perform
such a check against all available Higgs searches from LEP, the Tevatron, and
the LHC\@. This paper presents the upgrades to the code in \HBv{5} compared to
the previous version \HBv{4} described in Ref.~\cite{Bechtle:2013wla} and discusses the most important new features.

The general approach of \HB remains unchanged compared to \HBv{4} and we refer
to Ref.~\cite{Bechtle:2013wla} for the details. For each Higgs boson of the
investigated model, based on the model predictions input by the user, \HB
selects the most sensitive limit by comparing the model predictions to the
\emph{expected} limits of all analyses. It then checks the \emph{observed} limit
of this selected analysis against the model predictions to obtain a bound for
each Higgs boson. The model parameter point is considered allowed if none of the
Higgs bosons are excluded by the corresponding selected analysis. All of the
individual limits implemented in \HB are exclusion limits at $95\%$ confidence
level (C.L.). This procedure ensures that the overall combined limit is still
approximately at the \CL{95}\footnote{Applying all (or several) of the available
	analyses simultaneously would lead to a combined limit at a considerably lower
	confidence level than the original \CL{95} quoted for each analysis.} Likelihood
information has been made available by the experimental collaborations for
several analyses at LEP and at the LHC\@. \HB utilizes this detailed input to
reconstruct the corresponding \CL{95} limit in a nearly model-independent fashion or,
optionally, return the corresponding $\chi^2$ that can be used in a model fit.

In light of the increasing number of experimental search channels with improved
sensitivity one of the most important aspects of \HB is to provide an input
framework that works for a large class of BSM models and can incorporate all of
the required model predictions. Most of the major changes in \HBv{5} relate to
improvements in the input framework --- such as allowing additional cross
sections and branching ratios to be set as input or providing precise,
model-independent parametrizations of required input quantities. The extended
input framework is also used by the code
\HSv{2}~\cite{Bechtle:2013xfa,Bechtle:2014ewa, HS2Manual} which tests BSM models
against the Higgs rate measurements at the LHC (and the Tevatron).

In \cref{sec:theo} we describe the extended \HB input framework in detail,
discussing the available input schemes and the production and decay channels
supported by \HBv{5}. This includes the possibility of providing input beyond
the narrow width approximation as detailed in \cref{ssec:channelrates}. In
\cref{sec:exp} we review the experimental input used by \HB and consider
possible improvements to the input presentation that could make the experimental
results more readily useable. \Cref{sec:newHB} discusses functional changes in
\HBv{5}. This includes parametrizations of $VH$ and charged Higgs production
cross sections in the effective coupling approximation, as well as a description
of improvements in the derivation of exclusion limits from the available
likelihood information compared to the method first discussed in
Ref.~\cite{Bechtle:2015pma}. We give an overview of the technical changes
relevant to users of the code in \cref{sec:user} and conclude in
\cref{sec:summary}.

\section{Theoretical Input}
\label{sec:theo}

The input for \HB\ consists of the phenomenologically relevant physical quantities of
the Higgs sector, i.e.~the number of neutral and charged Higgs bosons that
should be considered, their masses, total decay widths, production and decay rates. By relying only on these physical quantities (in contrast to model-specific parameters), the code maintains a flexible input framework with rather minimal model assumptions.

\HBv{5} supports three types of input specified in the variable
\Code{whichinput} when initialising the code. These are the hadronic cross
section input (\Code{whichinput = 'hadr'}), the effective coupling input
(\Code{whichinput = 'effC'}), and SLHA (SUSY Les Houches
Accord~\cite{Skands:2003cj,Allanach:2008qq}) input (\Code{whichinput =
	'SLHA'}). The partonic input mode (\Code{whichinput = 'part'}) present in
previous versions of \HB has been removed and is no longer supported. We
describe the available methods of providing input to \HB in more detail in
Section~\ref{sec:user}.

For \Code{whichinput = 'hadr'} the inclusive hadronic\footnote{We use the term
	hadronic to distinguish from the partonic cross sections used in the
	deprecated partonic input mode. This includes the LEP cross sections,
	though they should be properly called leptonic cross sections.}
production cross sections have to be provided to the code. Cross
sections for various colliders and center-of-mass (CM) energies are
required --- namely LEP, Tevatron, and the LHC at \SIlist{7;8;13}{\TeV}.
If the considered production mode also exists in the SM, the input cross
section is normalized to the corresponding SM prediction for the same
Higgs mass, and otherwise specified in picobarn (\si{\pb}). In
\Code{effC} and \Code{SLHA} input the hadronic cross sections are
calculated internally from the provided effective couplings whenever
possible. The branching ratios (BRs) for all Higgs boson decays are also
required as input. In \Code{effC} mode the BRs for decay modes to SM
particles are per default approximated from the provided effective
couplings, and only the BRs for Higgs decay modes that are not present
for a SM Higgs boson have to be specified explicitly. In contrast, in
\Code{SLHA} input, the BRs for all decay modes are directly taken from
the SLHA \Code{DECAY} blocks.

\subsection{Neutral Higgs Bosons}
\label{sec:neutralinput}

\begin{table}[th!]
	\footnotesize
	\begin{tabularx}{\textwidth}{R X}
		\toprule
		CS_hj_ratio[j]        & SM normalized inclusive hadronic cross section for single Higgs production, $pp/p\bar{p}\to h_j$.
		This channel typically combines the $gg\to h_j$ and $pp/p\bar{p} \to b\bar{b}h_j$ channels that are given separately below.                                         \\
		CS_gg_hj_ratio[j]     & SM normalized inclusive hadronic cross section for the gluon fusion process, $pp/p\bar{p} \to gg \to h_j$. \NEW                             \\
		CS_bb_hj_ratio[j]     & SM normalized inclusive hadronic cross section for the $b\bar{b}$ associated Higgs production, $pp/p\bar{p} \to b\bar{b} h_j $. \NEW        \\
		CS_hjW_ratio[j]       & SM normalized inclusive hadronic cross section for Higgs production in association with a $W$ boson, $pp/p\bar{p} \to Wh_j$.                \\
		CS_hjZ_ratio[j]       & SM normalized inclusive hadronic cross section for Higgs production in association with a $Z$ boson, $pp/p\bar{p} \to Z h_j$.               \\
		CS_vbf_ratio[j]       & SM normalized inclusive hadronic cross section for the Higgs production in vector boson fusion, $pp/p\bar{p} \to q\bar{q} h_j $.            \\
		CS_tthj_ratio[j]      & SM normalized inclusive hadronic cross section for the $t\bar{t}$ associated Higgs production, $pp/p\bar{p} \to t\bar{t} h_j $.             \\
		CS_thj_tchan_ratio[j] & SM normalized hadronic cross section for single top quark associated Higgs production through $t$-channel exchange, $pp \to t q h_j $. \NEW \\
		CS_thj_schan_ratio[j] & SM normalized hadronic cross section for single top quark associated Higgs production through $s$-channel exchange, $pp \to t b h_j $. \NEW \\
		CS_tWhj_ratio[j]      & SM normalized hadronic cross section for Higgs production in association with a single top quark and a $W$ boson, $pp \to t W h_j$. \NEW    \\
		CS_qq_hjZ_ratio[j]    & SM normalized hadronic cross section for quark-intiated Higgs production in association with a $Z$ boson, $q\bar{q} \to Z \to Z h_j$. \NEW  \\
		CS_gg_hjZ_ratio[j]    & SM normalized hadronic cross section for gluon initiated Higgs production in association with a $Z$ boson, $gg \to Z h_j$. \NEW             \\
		CS_hjhi[j,i]          & Inclusive hadronic cross section for (non-resonant) double Higgs production, $pp/p\bar{p} \to h_j h_i$ (in pb). \NEW                        \\
		\bottomrule
	\end{tabularx}
	\caption{Hadronic cross section input for neutral Higgs bosons. The cross sections are inclusive in the electric charges of the  produced particles. Quantities
		added in \HBv{5} are labeled as \NEW.}
	\label{tab:hadrneutralHiggsinput}
\end{table}

The quantities needed to describe the production and decay rates of neutral
Higgs bosons are listed in
\cref{tab:hadrneutralHiggsinput,tab:lepneutralxs,tab:brneutral}. The hadronic
cross sections in \cref{tab:hadrneutralHiggsinput} have been extended by
separate input for gluon fusion and $b\bar{b}$ associated Higgs production,
whereas \HBv{4} only required the sum of the two processes (denoted as single
Higgs production). This is particularly relevant for exclusion limits from
analyses that have specific requirements on the $b$-jet multiplicity in the
event, as is e.g.~the case in searches for heavy BSM Higgs bosons decaying to
$\tau^+\tau^-$ (see also \cref{sec:tautau}). Furthermore, the cross sections for
processes of Higgs production in association with a single top quark have been
added. We distinguish between the $t$-channel process $q b\to t q h_j$
(specified in the 5-flavor scheme) and the $s$-channel process $qq' \to tbh_j$
(see Ref.~\cite{Demartin:2015uha} for a comprehensive discussion and for the NLO
QCD predictions in the SM). Similarly, separate cross sections for gluon- and
quark-initiated $Z$-boson associated Higgs production have been added to the \HB
input. These involve different Higgs couplings (see \cref{sec:VHprod}) and are
partly separated in the simplified template cross section (STXS) measurements
that are newly included in \HSv{2} (see Ref.~\cite{HS2Manual}). Since \HB
handles the input for \HS, this also means that this subchannel information is
available, such that differential information can be incorporated if it becomes
available in a framework similar to the STXS. Finally, the non-resonant double
Higgs production cross section has been added as input. Note that it is not
normalized to the SM prediction but should instead be given in~\si{\pb}.

The cross section input for LEP is unchanged with respect to \HBv{4}. For completeness, these
quantities are listed in \cref{tab:lepneutralxs}.

\begin{table}[t]
	\footnotesize
	\begin{tabularx}{\textwidth}{RX}
		\toprule
		CS_lep_hjZ_ratio[j]      & SM normalized LEP cross section for Higgs production in association with a $Z$ boson, $e^+e^- \to Z h_j$.           \\
		CS_lep_bbhj_ratio[j]     & SM normalized LEP cross section for the $b\bar{b}$ associated Higgs production, $e^+e^- \to b\bar{b} h_j $.         \\
		CS_lep_tautauhj_ratio[j] & SM normalized LEP cross section for the $\tau^+\tau^-$ associated Higgs production, $e^+e^- \to \tau^+\tau^- h_j $. \\
		CS_lep_hjhi_ratio[j,i]   & SM normalized LEP cross section for double Higgs production, $e^+e^- \to h_j h_i$.                                  \\
		\bottomrule
	\end{tabularx}
	\caption{LEP cross section input for neutral Higgs bosons. These are unchanged
		with respect to \HBv{4}.}
	\label{tab:lepneutralxs}
\end{table}

The branching ratio input for the decays of neutral Higgs bosons to SM particles
has been extended by Higgs decays into top quarks and flavor-changing leptonic
Higgs decays, see \cref{tab:brneutral}. Furthermore, we have generalized the BR
array for neutral Higgs boson decays to two neutral Higgs bosons, $h_k \to h_j
	h_i$, to allow for different Higgs bosons in the final states ($h_j \ne h_i$),
and added input BR arrays for neutral Higgs boson decays to a neutral Higgs
boson and a $Z$ boson, $h_j \to h_i Z$, and neutral Higgs boson decays to a
charged Higgs boson and a $W$ boson, $h_i \to H_i^\pm W^\mp$.

\begin{table}[t]
	\footnotesize
	\centering
	\begin{minipage}[b][][t]{0.45\linewidth}
		\begin{tabular}{Rl}
			\toprule
			BR_hjcc[j]     & $h_j\to c\bar{c}$      \\
			BR_hjss[j]     & $h_j\to s\bar{s}$      \\
			BR_hjtt[j]     & $h_j\to t\bar{t}$ \NEW \\
			BR_hjbb[j]     & $h_j\to b\bar{b}$      \\
			BR_hjmumu[j]   & $h_j\to \mu^+\mu^-$    \\
			BR_hjtautau[j] & $h_j\to \tau^+\tau^-$  \\
			BR_hjWW[j]     & $h_j\to W^+W^-$        \\
			BR_hjZZ[j]     & $h_j\to ZZ$            \\
			BR_hjgaga[j]   & $h_j\to \gamma\gamma$  \\
			BR_hjZga[j]    & $h_j\to Z\gamma$       \\
			BR_hjgg[j]     & $h_j\to gg$            \\
			\bottomrule
		\end{tabular}
	\end{minipage}\quad
	\begin{minipage}[b][][t]{0.45\linewidth}
		\begin{tabular}{Rl}
			\toprule
			BR_hjinvisible[j] & $h_j\to \text{invisible}$       \\
			BR_hkhjhi[k,j,i]  & $h_k\to h_j h_i$ \NEW           \\
			BR_hjhiZ[j,i]     & $h_j\to h_i Z$ \NEW             \\
			BR_hjemu[j]       & $h_j\to e^\pm \mu^\mp$ \NEW     \\
			BR_hjetau[j]      & $h_j\to e^\pm \tau^\mp$ \NEW    \\
			BR_hjmutau[j]     & $h_j\to \mu^\pm \tau^\mp $ \NEW \\
			BR_hjHpiW[j,i]    & $h_j\to H_i^\pm W^\mp$ \NEW     \\
			\bottomrule
		\end{tabular}
	\end{minipage}
	\caption{Branching ratios for neutral Higgs bosons. Possible decay modes for a
		SM-like Higgs bosons are on the left and decays involving new physics or flavor
		violation on the right. Quantities added in \HBv{5} are labeled as \NEW.}
	\label{tab:brneutral}
\end{table}

Instead of giving the hadronic and leptonic Higgs production cross sections and
branching fractions directly, \HB also features an effective coupling (or scale
factor) approximation for all these quantities. In case this approximation is
employed, the effective couplings listed in \cref{tab:effCneutral} have to be
provided. With respect to \HBv{4} we have changed the entire input from
\textit{squared}  effective couplings (or scale factors) to the \textit{sign-sensitive}
single effective couplings (or scale factors). This allows us to take into account
interference effects e.g.~in the prediction for the $h_j Z$ production cross
section. Furthermore, we removed the effective (squared) $h_j ggZ$ coupling
present in earlier versions. Instead, the $gg\to h_j Z$ contribution is derived
from the $h_j tt$ and $h_j bb$ effective couplings. Note that the loop-induced
$h_j gg$, $h_j \gamma\gamma$ and $h_j\gamma Z$ couplings are still free input
quantities not derived from the other coupling parameters.

\begin{table}[t]
	\footnotesize
	\begin{tabularx}{\textwidth}{RX}
		\toprule
		ghjcc_s[j], ghjcc_p[j]         & SM normalized effective Higgs couplings to charm quarks            \\
		ghjss_s[j], ghjss_p[j]         & SM normalized effective Higgs couplings to strange quarks          \\
		ghjtt_s[j], ghjtt_p[j]         & SM normalized effective Higgs couplings to top quarks              \\
		ghjbb_s[j], ghjbb_p[j]         & SM normalized effective Higgs couplings to bottom quarks           \\
		ghjmumu_s[j], ghjmumu_p[j]     & SM normalized effective Higgs couplings to muons                   \\
		ghjtautau_s[j], ghjtautau_p[j] & SM normalized effective Higgs couplings to tau leptons             \\
		ghjWW[j]                       & SM normalized effective Higgs coupling to $W$ bosons               \\
		ghjZZ[j]                       & SM normalized effective Higgs coupling to $Z$ bosons               \\
		ghjZga[j]                      & SM normalized effective Higgs coupling to a $Z$ boson and a photon \\
		ghjgaga[j]                     & SM normalized effective Higgs coupling to photons                  \\
		ghjgg[j]                       & SM normalized effective Higgs coupling to gluons                   \\
		ghjhiZ[j,i]                    & effective $h_jh_iZ$ coupling normalized to \cref{eq:hhZref}        \\
		\bottomrule
	\end{tabularx}
	\caption{Effective Higgs couplings for neutral Higgs bosons. The fermionic
		couplings have a CP-even scalar (\Code{_s}) and a CP-odd pseudoscalar
		(\Code{_p}) part. All of these are \NEW as they are now non-squared and
		sign-sensitive.}
	\label{tab:effCneutral}
\end{table}

The scalar and pseudoscalar components of the Higgs couplings to a generic
fermion pair $f\bar{f}$ are defined through
\begin{equation}
	g_{h_jf\bar{f}} = \iu (g_{s,\,h_jf\bar{f}} + g_{p,\,h_jf\bar{f}} \gamma_5)\eqcomma
	\label{eq:effC_fermions}
\end{equation}
where $g_s$ and $g_p$ are the real-valued scalar and pseudoscalar coupling
constants. As effective couplings, they are both normalized to the SM value of
$g_s$ for the corresponding fermion given by
\begin{equation}
	g_s^\text{ref}=g^\text{SM}_{hf\bar{f}}= \frac{e m_f}{2\sw M_W},
\end{equation}
with electric charge $e$, the weak mixing angle $\theta_\text{w}$, fermion mass $m_f$ and the $W$-boson mass $M_W$.
The couplings to $W$ and $Z$ bosons are normalized to the corresponding SM tree-level couplings
\begin{equation}
	g^\text{SM}_{hZZ} = \frac{e M_Z^2}{\sw M_W}\eqcomma\quad g^\text{SM}_{hWW} = \frac{e M_W}{\sw}\,,
	\label{eq:effC_bosons}
\end{equation}
where $M_Z$ is the mass of the $Z$-boson. The loop-induced effective couplings
to $\gamma\gamma$ and $Z\gamma$ are best defined through the partial decay
widths normalized to the SM-value for the same Higgs mass. This can also be used
for the $gg$ effective coupling, however it is a better approximation in this
case to use the normalized gluon fusion production cross section. Either way,
these yield the squared effective coupling whose square-root is the input
expected by \HBv{5}. The sign of these loop-induced couplings does not enter any
observables, so the positive square root can be used without loss of generality.

Finally, the $h_jh_iZ$ coupling does not have a SM equivalent that could be used
for normalization. It is instead normalized to
\begin{equation}
	g_{hh'Z}^\text{ref}=\frac{e}{4\sw\cw}\eqdot \label{eq:hhZref}
\end{equation}

More details on the effective coupling input and how it is used to approximate
the hadronic cross sections can be found in Ref.~\cite{Bechtle:2013wla}.

\subsection{Charged Higgs Bosons}
\label{ssec:charged}

The \HB input framework has been broadly extended in the charged Higgs sector.
We list all relevant charged Higgs sector quantities in
\cref{tab:chargedHiggsinput}. \HBv{5} supports direct charged Higgs boson
production at hadron colliders, including $H^\pm_j$ production in association
with a top or charm quark and a bottom quark as well as flavor-suppressed
production in association with lighter quark jets. We also include charged Higgs
production in association with a vector boson or a neutral Higgs boson, as well
as charged Higgs boson production in vector boson fusion and charged Higgs pair
production. Note that all hadronic cross sections are directly given in~pb, and
not specified as normalized quantities. All input cross sections are required to
be summed over the two possible charges. Note that at present there is no
effective coupling input for the charged Higgs bosons.

For light charged Higgs bosons with mass below the top quark mass, the most
important search channel is top quark pair production with successive decay of one top
quark to a charged Higgs boson and a bottom quark. \HB thus also requires the
branching fractions for $t\to H^+_j b$ and $t \to W^+ b$ --- where the latter is
needed to check for model assumptions --- as input. The charged Higgs branching
fractions have been extended by the decays to top and bottom quarks, $W$ and $Z$
bosons, as well as neutral Higgs and $W$ bosons.

Note that, thus far\footnote{As of June 2020.}, LHC searches have only
considered $pp\to H^\pm t b$ and $H^\pm$ production in vector boson fusion as
direct production channels. The remaining cross sections listed in the upper
section in \cref{tab:chargedHiggsinput} are therefore only placeholders at the
moment, and setting them to zero in a \HB\ run will not affect the results until
relevant experimental results become available.
\begin{table}[tb]
	\footnotesize
	\begin{tabularx}{\textwidth}{RX}
		\toprule
		CS_Hpmjtb[j]     & Hadronic cross section for $pp \to H_j^\pm t b$ production \NEW                             \\
		CS_Hpmjcb[j]     & Hadronic cross section for $pp \to H_j^\pm c b$ production \NEW                             \\
		CS_Hpmjbjet[j]   & Hadronic cross section for $pp \to H_j^\pm b + \text{light jet}$ production \NEW            \\
		CS_Hpmjcjet[j]   & Hadronic cross section for $pp \to H_j^\pm c + \text{light jet}$ production \NEW            \\
		CS_Hpmjjetjet[j] & Hadronic cross section for $pp \to H_j^\pm  + 2$ light jets production \NEW                 \\
		CS_HpmjW[j]      & Hadronic cross section for $pp \to H_j^\pm W^\mp$ production \NEW                           \\
		CS_HpmjZ[j]      & Hadronic cross section for $pp \to H_j^\pm Z$ production \NEW                               \\
		CS_vbf_Hpmj[j]   & Hadronic cross section for $pp \to H_j^\pm q\bar{q}$ production in vector boson fusion \NEW \\
		CS_HpjHmj[j]     & Hadronic cross section for $pp \to H_j^+ H_j^-$ production \NEW                             \\
		CS_Hpmjhi[j,i]   & Hadronic cross section for $pp \to H_j^\pm h_i$ production \NEW                             \\
		\midrule
		BR_tWpb[j]       & Branching ratio for the top quark decay $t\to W^+ b$                                        \\
		BR_tHpjb[j]      & Branching ratio for the top quark decay $t\to H_j^+ b$                                      \\
		\midrule
		BR_Hpjcs[j]      & Branching ratio for  $H_j^+\to c\bar{s}$                                                    \\
		BR_Hpjcb[j]      & Branching ratio for $H_j^+\to c\bar{b}$                                                     \\
		BR_Hpjtaunu[j]   & Branching ratio for $H_j^+\to \tau^+\nu_\tau$                                               \\
		BR_Hpjtb[j]      & Branching ratio for  $H_j^+\to tb$ \NEW                                                     \\
		BR_HpjWZ[j]      & Branching ratio for $H_j^+\to W^+ Z$ \NEW                                                   \\
		BR_HpjhiW[j,i]   & Branching ratio for $H_j^+\to h_i W^+$ \NEW                                                 \\
		\bottomrule
	\end{tabularx}
	\caption{Hadronic charged Higgs boson production cross sections (in pb), top quark branching ratios, and
		branching ratios for charged Higgs bosons. For the production cross sections the
		input has to be given for the sum of $H^+$ and $H^-$ production. Quantities
		added in \HBv{5} are labeled as \NEW.}
	\label{tab:chargedHiggsinput}
\end{table}

\subsection{Input Beyond the Narrow Width
	Approximation}\label{ssec:channelrates}

All of the input schemes described above rely on the narrow width
approximation (NWA) to construct the signal rates in specific collider
channels from the provided cross sections and branching ratios. As such, in
the NWA the \emph{channel~rate} $r^{p,d}$ is given by
\begin{equation}
	r^{p,d}\approx\sigma^p \cdot \text{BR}^d\eqcomma
\end{equation}
where $p$ denotes the production and $d$ the decay mode of the channel. In cases
where the NWA is not applicable, the hadron collider channel rates $r^{p,d}$ for
neutral Higgs bosons can be specified directly by the user, which then replace
the corresponding values obtained from the NWA\@. In this way, individual
channel rates can be set while keeping the remaining \HB input unchanged. For
instance, this is relevant if one of the neutral Higgs bosons of the model has a
very large width, while the narrow width approximation is applicable for the
remaining particles. Non-trivial modifications through signal-signal or
signal-background interference can also be accounted for by explicitly setting
the channel rate. For instance, destructive signal-signal interference of two
heavy Higgs bosons can appear in the MSSM with
CP-violation~\cite{Fuchs:2017wkq}, leading to sizable differences in the
exclusion obtained from BSM Higgs-to-$\tau^+\tau^-$ searches, as compared to the
naive, incoherent combination of the individual Higgs boson signal rates (see
\eg the discussion in~\cite{Bahl:2018zmf}).\footnote{To include interference
	effects of Higgs bosons in a specific channel in \HB, their combined signal rate
	has to be provided as channel rate in the user input of one of the Higgs bosons,
	while the channel rate of the other interfering Higgs boson(s) has to be set to
	zero in order to avoid double-counting.}

Note that there is currently no way to specify channel rates for charged Higgs
processes. Channel rates can be input using the \texttt{Fortran} subroutine
interface (using the subroutine
\Code{HiggsBounds_neutral_input_hadr_channelrates} or, for a single process, the
subroutine \Code{HiggsBounds_neutral_input_hadr_channelrates_single}), see the
online documentation for details. These subroutines expect the channel rates to
be normalized as $r^{p,d}/\sigma_\text{SM}^p$ where $\sigma_\text{SM}^p$ is the
corresponding production cross section for a SM-like Higgs boson of the same
mass. The \Code{HBwithchannelrates} example programs illustrates the use of
explicitly set channel rates.

Besides accounting for width effects in the theoretical input for the channel
rates, the experimental limits in various search channels are
often provided as a function of the total decay width. In that case, this
width-dependence of the limit is fully implemented in \HBv{5} and accounted for
in the model testing, irrespectively of whether the NWA was employed or the
channel rates have been set directly by the user.

\section{Experimental Input}
\label{sec:exp}
In this section we describe the experimental results that are used by \HB{}, and
we address possible limitations of the application of Higgs search limits to a
model. In this context we discuss how search limits (at fixed
confidence level (C.L.) or as a likelihood) should be presented, in particular
how they should be parametrized and what information is required in order to
apply an experimental limit to (nearly) arbitrary Higgs models. Finally, we make
suggestions for possible future refinements in the presentation of experimental
limits.

\subsection{Experimental data in \HB{} and Limitations of Applicability}

\HB currently incorporates results from LEP~\cite{Searches:2001aa,
	Searches:2001ab, Searches:2001ac, Abbiendi:2001kp, ALEPH:2002gcw,
	Abbiendi:2002qp, Abdallah:2003ry, Abbiendi:2003sc, Abdallah:2003wd,
	Achard:2004cf, Abdallah:2004wy, Schael:2006cr, Abbiendi:2007ac,
	Abbiendi:2008aa, Abbiendi:2013hk}, the Tevatron~\cite{Aaltonen:2008ec,
	ChristianSchwanenberger:2008ita, Abazov:2008wg, HaraldFox:2008yta,
	Abazov:2009aa, Aaltonen:2009ke, MaikoTakahashi:2009mxa, Aaltonen:2009vf,
	Abazov:2009yi, Abazov:2010ci, Aaltonen:2010cm, Abazov:2010ct,
	Benjamin:2010xb, FabriceCouderc:2010yca, Abazov:2010zk,
	TEVNPHWorking:2011aa, Abazov:2011ed, SubhenduChakrabarti:2011gfa,
	Abazov:2011jh, Abazov:2011qz, Benjamin:2011sv, TEVNPH:2012ab, D0:2012asa,
	J.-F.Grivaz:2012gba, D0:2012jsa, D0:2012msa, D0:2012osa, D0:2012psa,
	Aaltonen:2012qt, D0:2012rsa, D0:2012ssa, GuoChen:2012yga, Group:2012zca,
	CDFNotes, D0Notes}, and the ATLAS~\cite{ATLAS:2011aa, ATLAS:2011af,
	Aad:2011ec, ATLAS:2011jka, ATLAS:2012ac, ATLAS:2012ad, ATLAS:2012ae,
	ATLAS:2012aha, ATLAS:2012cpa, ATLAS:2012dsy, ATLAS:2012foa, ATLAS:2012kja,
	ATLAS:2012mja, ATLAS:2012nja, Aad:2012tfa, Aad:2012tj, ATLAS:2012toa,
	ATLAS:2012znl, ATLAS:2013nma, ATLAS:2013qma, ATLAS:2013wla, Aad:2014fia,
	Aad:2014iia, Aad:2014ioa, Aad:2014vgg, Aad:2014xva, Aad:2014yja,
	ATLAS:2014zha, Aad:2015agg, Aad:2015bua, ATLAS:2015dka, Aad:2015kna,
	Malone:2015mia, Aad:2015nfa, Aad:2015wra, Aad:2015xja, ATLAS:2016bza,
	ATLAS:2016ixk, ATLAS:2016kjy, ATLAS:2016lri, ATLAS:2016npe, Aaboud:2016okv,
	ATLAS:2016oum, Aaboud:2016oyb, ATLAS:2016yqq, Aaboud:2017fgj,
	Aaboud:2017gsl, Aaboud:2017rel, Aaboud:2017sjh, Aaboud:2017yyg,
	Aaboud:2018bun, Aaboud:2018cwk, Aaboud:2018eoy, Aaboud:2018esj,
	Aaboud:2018ewm, Aaboud:2018gjj, Aaboud:2018iil, Aaboud:2018ksn,
	Aaboud:2018sfi, Aaboud:2018sfw, ATLAS:2018xad, Aad:2019ojw, Aaboud:2019rtt,
	Aaboud:2019sgt, Aad:2019ugc, Aad:2019uzh, Aad:2019zwb} and
CMS~\cite{CMS:2012bea, Chatrchyan:2012dg, Chatrchyan:2012ft,
	Chatrchyan:2012sn, Chatrchyan:2012tx, CMS:2012yta, CMS:2012ywa, CMS:2013jda,
	Chatrchyan:2013mxa, Chatrchyan:2013vaa, CMS:2013yea, Chatrchyan:2013zna,
	Khachatryan:2014ira, Chatrchyan:2014tja, Khachatryan:2015cwa, CMS:2015iga,
	Khachatryan:2015lba, CMS:2015lza, CMS:2015mca, Khachatryan:2015nba,
	CMS:2015ocq, Khachatryan:2015qba, Khachatryan:2015qxa, Khachatryan:2015tha,
	Khachatryan:2015tra, Khachatryan:2015uua, Khachatryan:2015wka,
	Khachatryan:2015yea, Khachatryan:2016are, CMS:2016ilx, CMS:2016ncz,
	Khachatryan:2016sey, CMS:2016szv, CMS:2016tgd, CMS:2016tlj, CMS:2016vpz,
	Sirunyan:2017djm, Sirunyan:2017guj, Khachatryan:2017mnf, CMS:2017sbi,
	Sirunyan:2017uvf, Sirunyan:2018aui, Sirunyan:2018dvm, Sirunyan:2018mbx,
	Sirunyan:2018mot, Sirunyan:2018owy, Sirunyan:2018pzn, Sirunyan:2018qlb,
	Sirunyan:2018taj, Sirunyan:2018two, Sirunyan:2018zut, Sirunyan:2019bgz,
	Sirunyan:2019gou, Sirunyan:2019hkq, Sirunyan:2019pqw, Sirunyan:2019shc,
	CMS:2019sxh, Sirunyan:2019tkw, CMS:2019vgr, Sirunyan:2019wph,
	Sirunyan:2019wrn, Sirunyan:2019xls, Sirunyan:2020hwv, CMS:aya, CMS:bxa,
	CMS:lng, CMS:wxa, CMS:xxa, CMS:zwa} experiments at the LHC\@. A detailed
list of the implemented analyses is returned by the \Code{AllAnalyses}
executable (see \cref{ssec:commandline}). An up-to-date version of this
list, together with a bibliography of all implemented results is also
available on the webpage, and the \textsf{InspireHEP} cite keys of the
analyses are included in the \HB output (in the \Code{Keys.dat} file). We
expect all users of \HB to cite the relevant experimental analyses.

The application of the experimental exclusion limits to a model parameter point
is described in detail in Ref.~\cite{Bechtle:2013wla}. The basic procedure is as
follows: In the first step, based on the \emph{expected} exclusion limit (at
$\CL{95}$), \HB selects the most sensitive analysis for each Higgs boson of the
model. In the second step, the model predictions for each Higgs boson are
compared with the \emph{observed} limit from the particular experimental search
that is most sensitive to it. If the predicted signal rate exceeds the observed
limit for any of the Higgs bosons, the model parameter point is regarded as
excluded (at $\CL{95}$). The validity of this test depends, in short, on the
following basic assumptions (see Ref.~\cite{Bechtle:2013wla} for details):
\begin{itemize}
	\item the narrow width approximation is valid, \ie the signal rate can be
	      approximated by the product of the Higgs boson production cross
	      section and branching ratio\footnote{As  described in
		      \cref{ssec:channelrates}, exceptions to this assumption are possible
		      in specific cases.};
	\item background processes in the experimental analyses are not altered
	      significantly by the signal  (new physics) model;
	\item the kinematics of the signal processes are not altered significantly
	      with respect to the signal hypothesis employed in the experimental analysis
	      (typically, a scalar or pseudoscalar boson with renormalizable couplings).
\end{itemize}

Furthermore, there are experimental analyses that combine different Higgs boson
search channels. Such combined limits may require an applicability test, \ie a
check whether the parameter point fulfills the assumptions on which the
combination is based. If this applicability test fails, the corresponding search
limit is not considered in the \HB test of the parameter point. Examples of such
analyses are searches for a SM-like Higgs boson or an invisibly decaying Higgs
boson, where various production modes are combined under the assumption of a
signal composition as predicted in the SM\@. If no further information on the
signal efficiencies is given (see below), the application of these search limits
to a model requires a  \emph{SM-likeness test} of the parameter point (see
Ref.~\cite{Bechtle:2013wla} for details). In short, this test checks whether the
model-predicted signal composition is similar to the SM prediction, where all
relevant search channels quoted in the analysis are taken into account, and
their inclusive cross sections are used as weights in the determination of the
maximally allowed deviation of the individual channel signal strength from the
total signal strength. The details of the SM-likeness test procedure are
described in Ref.~\cite{Bechtle:2013wla} and have not been changed in \HBv{5}.
After the LHC discovery of a SM-like Higgs boson, the focus of Higgs searches
has somewhat shifted towards more model-independent, less combined search
channels. Nevertheless, there are still many searches that combine different
production or decay modes, assuming the relative contributions to be equal to
those predicted in the SM, as \eg motivated by predictions in pure scalar
singlet extensions of the SM with a non-zero singlet-doublet mixing (see \eg
Refs.~\cite{Robens:2015gla, Robens:2016xkb, Lewis:2017dme, Ilnicka:2018def,
	Costa:2015llh, Muhlleitner:2017dkd, Dawson:2017jja, Robens:2019kga} for
phenomenological studies in the LHC Run-2 era).

The SM-likeness test would not be needed if more information was provided
publicly for the considered experimental analysis. The limit is typically
reported on an inclusive cross section, $\sigma_\text{tot}$, often also as
signal strength, $\mu$, \ie normalized to the corresponding SM
prediction.\footnote{In special cases where no signal rate limit can be
	constructed \HB can also implement limits on other quantities that rely on
	additional model assumptions. This is currently the case for the CMS
	$gg\to\phi\to t\bar{t}$ search~\cite{Sirunyan:2019wph} that constrains the
	effective $\phi t\bar{t}$ coupling.} For a combination of search channels
$i=1,\dots,N$, this signal strength can be calculated as
\begin{align}
	\mu = \frac{\sum_i \epsilon_i \sigma_i}{\sum_i \epsilon_i^\text{SM} \sigma_i^\text{SM}},
\end{align}
where $\sigma_i$ ($\sigma_i^\text{SM}$) denotes the inclusive signal rate for
search channel $i$ --- comprised of one production and one decay mode --- in the
model (SM), respectively, and $\epsilon_i$ ($\epsilon_i^\text{SM}$) is the
signal efficiency of channel $i$ in the analysis, i.e.~the fraction of signal
events that pass the event selection, as predicted in the model (SM). If the
three basic assumptions listed in the bullet points above are fulfilled, we have
--- to a good approximation --- $\epsilon_i = \epsilon_i^\text{SM}$. A
complication arises, however, if the experimental analysis does not provide
information of the signal efficiencies of the involved signal channels as
predicted in the SM, $\epsilon_i^\text{SM}$. Unfortunately, it has been common
practice to \emph{not} release this information until now.\footnote{Recently,
	ATLAS released signal efficiency information in one of their analyses, see
	Ref.~\cite{ATLAS:2018xad}.} If the $\epsilon_i^\text{SM}$ are unknown, we can
only safely calculate $\mu$ in the model if $\sigma_i/\sigma_i^\text{SM} \approx
	\mu$ for all channels $i=1,\dots, N$, and this is exactly what the
\emph{SM-likeness test} in \HB verifies.

Besides the usual \CL{95} limits, \HB{} contains additional exclusion likelihood
approximations for several cases. These are available for the main Higgs boson
search channels as well as the SM and MSSM Higgs boson search combinations at
LEP~\cite{Schael:2006cr,Bechtle:2013wla}. Moreover, \HB reconstructs the
exclusion likelihood from BSM Higgs boson searches in the $\tau^+\tau^-$ final
state by ATLAS~\cite{Aaboud:2017sjh,Aad:2020zxo} and
CMS~\cite{CMS:2015mca,Sirunyan:2018zut}, based on the numerical results
presented in a single narrow resonance parametrization with two production
modes. More details will be given in \cref{sec:tautau}.

\subsection{Recommendations for the Presentation of Future Search Results}

Recently, a joint effort within the experimental and theoretical community led
to the release of recommendations for the presentation of experimental
results~\cite{Abdallah:2020pec}. Following and partly expanding upon these
recommendations, we propose the following guidelines for the publication of
limits for LHC searches for new scalar bosons:
\begin{enumerate}
	\item upper limits on the cross sections of the signal processes should be
	      presented as a function of \emph{all relevant kinematical parameters},
	      \eg the masses and total widths of the involved scalar boson(s);
	\item the search results should always contain the expected \emph{and} the
	      observed limit;
	\item if the signal is comprised of several signal channels
	      (\ie different production and/or decay modes), the limit is set on a
	      common scale factor --- the signal strength $\mu$ --- or a total signal
	      rate. In this case, the \emph{signal efficiency} of each signal
	      channel should be provided as a function of \emph{all relevant
		      kinematical parameters} (see point 1);
	\item if the limit is presented as a normalized signal rate (\eg to the
	      SM prediction), the \emph{reference signal rate} should be quoted by
	      the experimental analysis along with the result, thus enabling the
	      recalculation of the limit on the signal rate's absolute value;
	\item the search limit should always be presented at \CL{95};
	\item in addition, it would be beneficial to present results as \emph{exclusion
		      likelihoods}, using the same parametrization as the one used for the
	      \CL{95} upper limit (see point 1).
\end{enumerate}

These guidelines should result in a format of the search limit that is to a
large extent model-independent, in the sense that all dependences on kinematic
parameters are fully described and can thus be incorporated in a recast of the
limit onto specific models. Presenting both the expected and observed result
enables a well-defined selection of the most sensitive analysis out of many
search results --- as already done in \HB --- such that the derived global
exclusion result can be interpreted at the \CL{95}. As already mentioned in the
previous subsection, quoting the signal efficiencies is necessary for the proper
determination of the signal strength in the model if several signal channels are
involved in the signal process, and would be a better alternative to the
\emph{SM-likeness test} that needs to be applied otherwise.

As already mentioned, the BSM Higgs searches in the $\tau^+\tau^-$ final state
by CMS~\cite{CMS:2015mca,Sirunyan:2018zut} and ATLAS~\cite{Aad:2020zxo,Aaboud:2017sjh} pioneered the publication of
(multi-dimensional) exclusion likelihoods in addition to the usual $\CL{95}$
upper limits at the LHC\@. These likelihoods were presented for a simplified
model of a single narrow scalar resonance $\phi$, parametrized in terms of its
mass, $m_\phi$, the signal rate for single scalar production, $\sigma(pp\to \phi
	\to \tau^+\tau^-)$, and the signal rate for scalar production in association
with bottom-quarks, $\sigma(pp\to b\bar{b}\phi \to \tau^+\tau^-)$ (see
\cref{sec:tautau} for details). This likelihood information has already turned
out to be very useful in various phenomenological
analyses (see e.g.~Refs.~\cite{Bechtle:2016kui, Bagnaschi:2017tru, Costa:2017gup,
	Bagnaschi:2018zwg}). Therefore, we strongly encourage the publication of
exclusion likelihoods (in a similar form) also for other BSM Higgs search
channels. We list in \cref{tab:likelihood-wishlist} a set of search channels,
along with the relevant kinematic and signal rate parameters, which we deem
suitable for providing this public information.

The table furthermore lists some model candidates for which these search results
would be useful. In particular, fermionic final states are most relevant in
models with additional Higgs doublets such as the MSSM or the 2HDM, where the
couplings of $\phi$ to fermions can be large even in the alignment limit where
the \SI{125}{\GeV} Higgs boson, denoted as $h_{125}$, has SM-like
couplings~\cite{Gunion:2002zf, Craig:2013hca, Carena:2013ooa, Carena:2014nza,
	Bechtle:2016kui, Haber:2017erd, Dev:2014yca, Pilaftsis:2016erj,
	Benakli:2018ldd}. On the other hand, the highly sensitive $VV$ final states are
very important for singlet extensions, where the production rates of additional
BSM Higgs boson(s) are mixing-suppressed when requiring $h_{125}$ to be
approximately SM-like, see \eg Refs.~\cite{Robens:2015gla, Robens:2016xkb,
	Robens:2019kga} for recent discussions of the impact of these searches on the
model parameter space. The $pp\to\phi_2\to Z\phi_1$ process ($\phi_{1,2}\neq
	h_{125}$) is of particular interest in 2HDMs, where it is correlated with a
strong first order electroweak phase transition (see \eg
Refs~\cite{Dorsch:2014qja,Basler:2016obg,Dorsch:2017nza}). Resonant di-Higgs
signatures are especially prominent in singlet extensions, where the coupling of
the non-$h_{125}$ scalars to all SM particles are suppressed. In non-minimal
singlet extensions even resonant di-Higgs processes involving two additional
scalars and $h_{125}$ are well motivated~\cite{Robens:2019kga}. Searches for
charged Higgs bosons can also be highly complementary to neutral Higgs searches
if additional scalar doublets are present in the model. Note that this table is
neither complete, nor does it give a ranking in priority. Providing likelihood
information along with the usual \CL{95} limits is useful in \emph{any}
analysis, and should be done if feasible.

\begin{table}[tb]
	\footnotesize
	\begin{tabularx}{\textwidth}{lll}
		\toprule
		Search Channel                                           & Possible Relevant Parameters                                                                  & Model Motivation, \eg \\
		\midrule
		$pp \to \phi (+b\text{-jets}) $, $\phi \to \tau^+\tau^-$ & $M_\phi$, $\sigma(pp\to \phi \to \tau^+\tau^-)$, $\sigma(pp\to b\bar{b}\phi\to \tau^+\tau^-)$ & MSSM, 2HDM            \\
		$pp \to b\bar{b} \phi $, $\phi \to b\bar{b}$             & $M_\phi$, $\sigma(pp\to \phi\to b\bar{b})$, $\sigma(pp\to b\bar{b}\phi \to b\bar{b})$         & MSSM, 2HDM            \\
		$pp \to \phi \to t\bar{t} $                              & $M_\phi$, $g_{s,\phi t\bar{t}}$, $g_{p,\phi t\bar{t}}$, $\Gamma_\text{tot}$                   & MSSM, 2HDM            \\
		$pp \to \phi \to VV (V =Z, W^\pm)$                       & $M_\phi$, $\mu_{pp\to \phi}$, $\mu_{\text{VBF},V\phi}$, [$\Gamma_\text{tot}$]                 & singlet extensions    \\
		$pp \to \phi_2 \to Z \phi_1$                             & $M_{\phi_1}$, $M_{\phi_2}$, $\sigma(pp\to\phi_2\to Z\phi_1)$                                  & 2HDM                  \\
		$pp \to \phi \to h_{125}h_{125}$                         & $M_\phi$, $\sigma(pp\to \phi\to h_{125}h_{125})$, [$\Gamma_\text{tot}$]                       & singlet extensions    \\
		$pp \to \phi_2 \to h_{125} \phi_1$                       & $M_{\phi_1}$, $M_{\phi_2}$, $\sigma(pp\to\phi_2\to h_{125}\phi_1)$                            & singlet extensions    \\
		$pp \to tb \phi^\pm$, $\phi^\pm \to tb$                  & $M_{\phi^\pm}$, $\sigma(pp\to tb \phi^\pm \to tbtb)$                                          & MSSM, 2HDM            \\
		$pp \to tb \phi^\pm$, $\phi^\pm \to \tau\nu$             & $M_{\phi^\pm}$, $\sigma(pp\to tb \phi^\pm \to tb\tau\nu)$                                     & MSSM, 2HDM            \\
		\dots                                                    & \dots                                                                                         & \dots                 \\
		\bottomrule
	\end{tabularx}
	\caption{A wishlist for the publication of exclusion likelihoods in several Higgs search channels. The left column denotes the search channel, the middle column the possible relevant parameters (quantities in square brackets seem less important), and the right column lists examples of BSM models for which the result would be valuable. }
	\label{tab:likelihood-wishlist}
\end{table}

We also appreciate the efforts of using simplified workspaces provided by the
experimental collaborations. In this approach it might be
possible to retain the dominant theory nuisance parameters in the likelihood,
while all experimental nuisance parameters that are not correlated with the
dominant theory nuisance parameters are marginalized. Extensive tests would be
necessary to explore the feasibility of this approach and the potential gain in
information and precision over the current approach. A simplified approach based
on the JSON format~\cite{ATLAS:2019oik} has been proposed where only the
most relevant theoretical (and, if necessary, experimental) systematic
uncertainties are retained separately, and all other uncertainties are combined.

BSM Higgs boson searches at the LHC have so far considered \emph{inclusive}
signal processes, or, at least, have presented the result as a limit on the
inclusive cross section. In the future a possible new path in the presentation
of search results could be the  presentation of limits on signal rates in
specific phase space regions --- so-called fiducial signal rates --- instead of
unfolding the result onto the inclusive signal rate. This is analogous to
measurements of the discovered Higgs boson's signal rates, where strong efforts
have recently been made to define specific phase-space regions for measurements
in order to reduce the theory-dependence introduced in the unfolding process. In
a similar way, one could think of a generalization of the Simplified Template
Cross Section (STXS) framework~\cite{deFlorian:2016spz} to the case of search limits. As \HB is used as a
framework for \HS, which already incorporates STXS measurements of the Higgs
boson signal, it would be straight-forward to implement corresponding
phase-space-dependent limits also in \HB.

Lastly, it would be desirable in the future to improve and automize the
implementation of experimental search results in \HB. The first step is that the
experimental collaborations provide search limits in a machine-readable format,
\eg via their TWiki pages or via \texttt{HEPData}~\cite{Maguire:2017ypu}. This
is already done in many cases. As a next step, it would be useful to define a
common data format that contains all necessary information about the search
limit. Such data files can then be read in automatically by \HB. Such a data
interface would also allow the \HB user to select specific search limits for
their study in a very versatile way.

\section{New Features in \HBv{5}}
\label{sec:newHB}

The most important improvement in \HBv{5} over its predecessor $\HBv{4}$ is the
inclusion of experimental results from the \SI{13}{\TeV} LHC\@. However, we will
not discuss the newly implemented search limits in detail.
Instead, we refer to the output of the \Code{AllAnalyses} executable (see
\cref{ssec:commandline}) to get a complete list of the experimental results
included in a specific version of \HB. Instead, this section will describe the
most relevant functional changes within \HBv{5}.

\subsection{Effective Coupling Approximation for \boldmath{$\phi_i V$} Production}
\label{sec:VHprod}

In previous versions of \HB the cross section ratios for neutral Higgs
boson ($\phi$) production in association with a massive gauge boson,
$pp/p\bar{p} \to V\phi $ (with $V=W^\pm,Z$), were obtained in the effective
coupling approximation purely from the effective $\phi VV$ coupling. In \HBv{5}
we have extended this approximation \emph{beyond} the leading order by
including contributions proportional to the scalar and pseudo-scalar Higgs
couplings to top and bottom quarks, as well as interference effects.

The production of a neutral scalar boson $\phi$ in association with $W$-bosons
always (up to next-to-leading order in QCD) takes place via Higgs-strahlung and is necessarily
dependent on the coupling $g_{\phi WW}$ of the respective particle to the $W$
bosons. At next-to-next-to-leading order (NNLO) in QCD, corrections from virtual
top-quark loops arise which depend on the scalar coupling to top-quarks,
$g_{s,\,\phi t\bar{t}}$. In case of $Z\phi $ production, additional important
box-diagrams from the partonic process $g g \rightarrow Z\phi $ need to be
accounted for.

In terms of the effective couplings for a scalar particle $\phi$,
\cref{eq:effC_fermions} and \cref{eq:effC_bosons}, we define
\begin{equation}
	\begin{aligned}
		         &                                                                           & \kappa_W                                                                        & := \left(\frac{g_{\phi WW}}{g_{HWW}^{\text{SM}}} \right),
		         & \kappa_Z                                                                  & := \left(\frac{g_{\phi ZZ}}{g_{HZZ}^{\text{SM}}} \right),                       &                                                           & \\
		\kappa_t & := \left(\frac{g_{s,\,\phi t\bar{t}}}{g_{Ht\bar{t}}^{\text{SM}}} \right),
		         & \kappa_b                                                                  & := \left(\frac{g_{s,\,\phi b\bar{b}}}{g_{Hb\bar{b}}^{\text{SM}}} \right),
		         & \kappa_{\tilde{t}}                                                        & := \left(\frac{g_{p,\,\phi t\bar{t}}}{g_{Ht\bar{t}}^{\text{SM}}} \right),
		         & \kappa_{\tilde{b}}                                                        & := \left(\frac{g_{p,\,\phi b\bar{b}}}{g_{Hb\bar{b}}^{\text{SM}}} \right) \eqdot
	\end{aligned}
\end{equation}
The cross sections can been expanded as follows:
\begin{align}
	\sigma^{W\phi }[m_{\phi}] & \approx \kappa_W^2 \bar{\sigma}^{ W\phi}_{WW}[m_{\phi}] + 2 \kappa_W \kappa_t \bar{\sigma}^{W\phi }_{W t}[m_{\phi}], \label{eq:sigmawh}                            \\
	\sigma^{ Z\phi}[m_{\phi}] & \approx \sum_{\mathrlap{\hspace{-0.8em}a,b \in \{Z, t, b, \tilde t, \tilde b\}}}\kappa_a \kappa_b \bar{\sigma}^{ Z\phi}_{a b}[m_{\phi}]  \eqdot \label{eq:sigmazh}
\end{align}
Note that $W\phi$ production is largely dominated by the leading-order
Higgs-strahlung process and only gets minor corrections from virtual top-quark
loops. This is why the effect of bottom quarks has been neglected and only the
$\kappa_W \kappa_t$ interference term is considered in \cref{eq:sigmawh}. In
contrast, for $Z\phi$ production all possible combinations $\kappa_a \kappa_b$
have been included. For this expansion we neglect the effects from other
possible  scalar bosons in the model that may contribute due to non-vanishing
$\phi \phi' Z$ couplings. If these contributions turn out to be relevant in the
investigated model, we advise the user to directly provide the hadronic cross
sections instead of using the effective coupling approximation.

We calculate the inclusive $W \phi$ and $Z \phi$ production cross sections with
\vhatnnlotwo~\cite{Brein:2012ne,Harlander:2018yio} at NNLO in QCD\@. The
mass-dependent expansion coefficients $\bar\sigma^{V\phi}_{ab}[m_{\phi}]$ are
determined by using the CP-violating 2HDM implementation of \vhatnnlo{} to
calculate various cross sections for different combinations of the effective
couplings and solving the resulting system of linear equations. They are
symmetric under $a \leftrightarrow b$.

\vhatnnlo\ does not evaluate contributions from $b \bar b \to Z\phi $ for
CP-mixed scalars which could lead to sizable differences in scenarios with large
$\kappa_{\tilde b}$. However, within our approximation $g_{\phi\phi'Z} = 0$ at
tree-level
\begin{align}
	\frac{\sigma(b \bar b \rightarrow H Z)}{\kappa_b^2} = \frac{\sigma(b \bar b \rightarrow A Z)}{\kappa_{\tilde b}^2}
\end{align}
holds for a pure CP-even scalar $H$ and CP-odd scalar $A$ with respective
effective couplings $\kappa_b$ and $\kappa_{\tilde{b}}$ (and equal masses). We
therefore use the \vhatnnlo{} SM Higgs boson implementation to determine the $b
	\bar b \rightarrow H Z$ contribution, and consider it for both the
$\bar{\sigma}^{Z\phi }_{bb}$ and $\bar{\sigma}^{Z\phi}_{\tilde b \tilde b}$ term
in \cref{eq:sigmazh}.

The cross section calculation for a pure CP-even scalar particle $H$ in
\vhatnnlo{} is more precise than the calculation for a CP-mixed scalar
boson $\phi$. For a SM-like CP-even Higgs boson \cref{eq:sigmazh} therefore produces
a less accurate result than the dedicated calculation for a pure CP-even scalar boson in \vhatnnlo.
To circumvent this problem, we apply a $K$-factor approach and rescale the cross
section for the CP-mixed scalar $\phi$ by the ratio of the more accurate
CP-even calculation $\sigma^{ZH}$ and the scalar terms of the CP-mixed
calculation $\sigma^{\phi Z}$, \ie
\begin{align}
	\sigma^{\phi Z}[m_{\phi}]  \approx \Big(\sum_{\mathrlap{\hspace{-0.8em}a,b \in \{Z, t, b, \tilde t, \tilde b\}}} \kappa_a \kappa_b \bar{\sigma}^{\phi Z}_{ab}[m_{\phi}]\Big) \cdot K(\kappa_t, \kappa_b, \kappa_Z, m_{\phi})\eqdot \label{eq:vhkfac1}
\end{align}
with
\begin{align}
	K(\kappa_t, \kappa_b, \kappa_Z, m_{\phi}) \equiv \frac{\sum\limits_{\mathrlap{\hspace{-0.8em}a,b \in \{Z, t, b\}}} \kappa_a \kappa_b \bar{\sigma}^{H Z}_{ab}[m_{\phi}]}{\sum\limits_{\mathrlap{\hspace{-0.8em}a,b \in \{Z, t, b\}}} \kappa_a \kappa_b \bar{\sigma}^{\phi Z}_{ab}[m_{\phi}]}\eqdot \label{eq:vhkfac2}
\end{align}
The definition in \cref{eq:vhkfac1} ensures that for a pure scalar, i.e.~for
$\kappa_{\tilde t} = \kappa_{\tilde b} = 0$, our approximation coincides with
the more accurate SM-like Higgs boson calculation in \vhatnnlo. For a SM-like
Higgs boson with all $\kappa_i = 1$, the $K$-factor in \cref{eq:vhkfac2} ranges
between 1 and 2, with values larger than 1.1 only appearing for $m_\phi \gtrsim
	\SI{400}{\GeV}$. For gauge-phobic particles with $\kappa_Z=0$, $K$ is typically
of the order of 2 and can range up to 5 for masses below \SI{10}{\GeV}.

\begin{figure}
	\centering
	\includegraphics[width=0.6\textwidth]{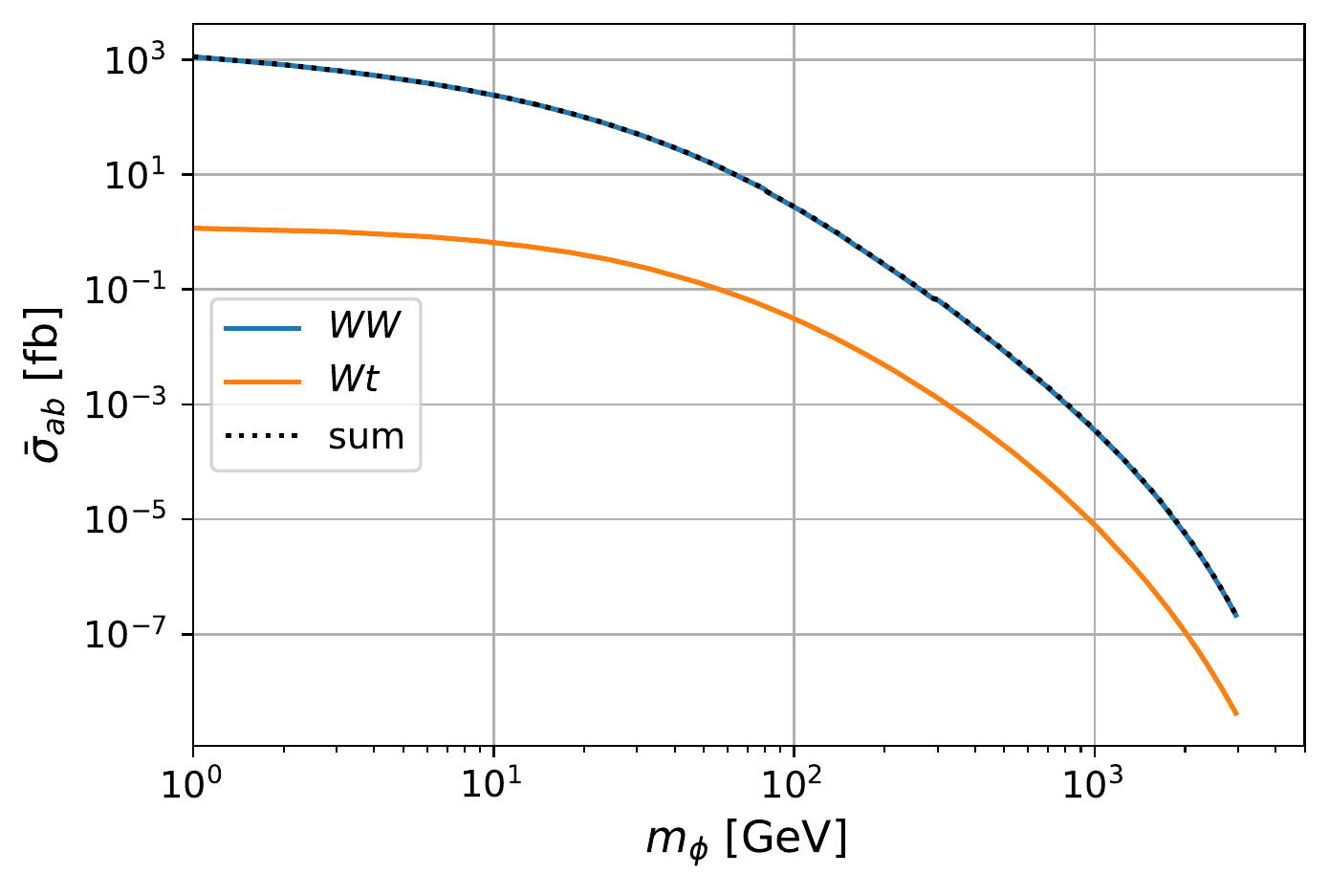}
	\caption{Cross section contributions of \cref{eq:sigmawh} as a function of
		$m_\phi$ for the process $pp\to W\phi$ at the LHC with a CM energy of
		$\SI{13}{\TeV}$. The dotted black line indicates the sum of the
		individual contributions.}
	\label{fig:XSWH}
\end{figure}

\begin{figure}
	\centering
	\includegraphics[width=\textwidth]{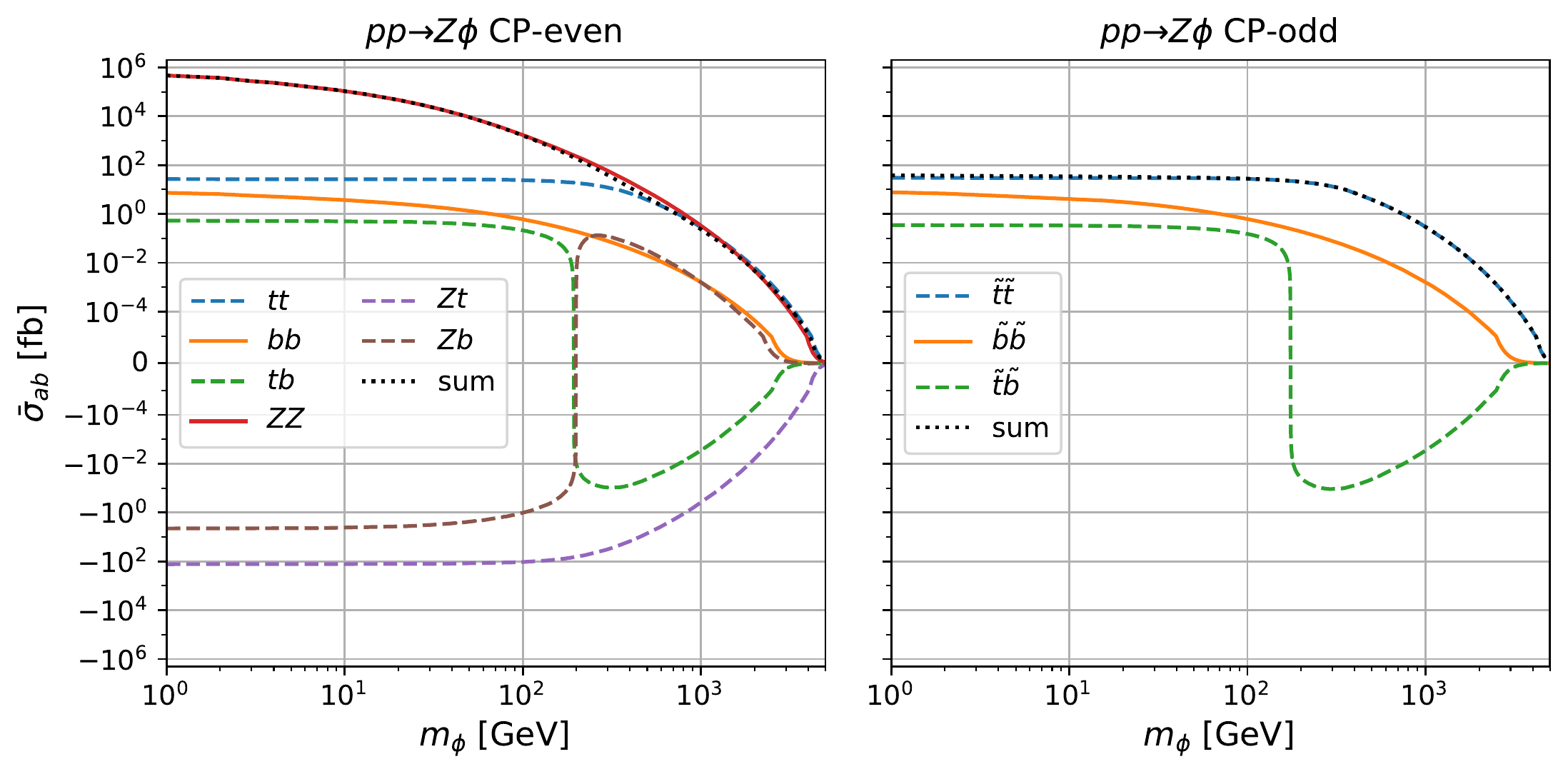}
	\caption{Cross section contributions of \cref{eq:sigmazh} as a function of $m_\phi$ for
		the process $pp\to Z\phi$ at the LHC with a CM energy of
		$\SI{13}{\TeV}$. The dashed lines indicate contributions originating
		entirely from the loop-induced $gg\to Z\phi$ subprocess. The dotted
		black lines indicate the sum of the individual contributions.}
	\label{fig:XSZH}
\end{figure}

The coefficients $\bar{\sigma}_{ab}$ in
\cref{eq:sigmawh,eq:sigmazh} are shown in \cref{fig:XSWH,fig:XSZH} as a function of $m_\phi$. \Cref{fig:XSWH}
shows the two contributions to the $pp\to W\phi$ process. \Cref{fig:XSZH} shows
the contributions to $pp\to Z\phi$ grouped into pure CP-even (left) and pure
CP-odd (right) contribtions. The dominant CP-even contribution is
$\bar{\sigma}_{ZZ}$ until the $tt$ and $Zt$ contributions become similarly
relevant for $m_\phi\gtrsim \SI{500}{\GeV}$. The CP-odd $\tilde{t}\tilde{t}$
contribution is nearly identical to the $tt$ contribution and is the largest
contribution for the CP-odd case, where $ZZ$ contributions are absent. Note that
the $bb$ and $\tilde{b}\tilde{b}$ contributions contain the $b\bar{b}\to Z\phi$
subprocess in addition to the $b$-quark boxes of the $gg\to Z\phi$ subprocess.
All cross-terms proportional to the product of a CP-even and a CP-odd
coupling vanish in our parametrization. These can only originate from the
contribution of additional Higgs bosons, which is neglected here.

Note that our expansions in \cref{eq:sigmawh,eq:sigmazh} have the following
limitations in their applicability to BSM Higgs models:
\begin{itemize}
	\item In all processes, the higher-order correction terms only account for
	      virtual top- and bottom-quark loops, thus we assume that no other
	      particles with similar quantum numbers run in the loop and give a
	      significant contribution. Models with additional, color-charged
	      particles with scalar interactions, as \eg the scalar top squarks in
	      supersymmetric theories, may require a proper calculation of the
	      quantum corrections of the additional particles.
	\item As scalar-scalar-vector interactions, $g_{\phi\phi'V}^{\text{model}}$, are
	      not accounted for, care must be taken in models with several light scalar
	      degrees of freedom with non-negligible interaction between these particles.
	      These may contribute in $s$-channel diagrams with the additional scalar
	      particles as propagators.
\end{itemize}
The approximation described here is a very significant improvement over the effective
coupling approximation for $pp\to VH$ in previous versions of \HB. It is
automatically used in the effective coupling input. In models where the
assumptions above are satisfied it can also be used to substitute an explicit
calculation of $\sigma(pp\to VH)$ for the hadronic input scheme. In this case,
the approximated hadronic cross sections can be accessed through the functions
in the \Code{access_effC.f90} file. Furthermore, this approximation is not only
used for the inclusive $\sigma(pp\to ZH)$ cross section but also provides
separate $\sigma(qq\to ZH)$ and $\sigma(gg\to ZH)$ cross sections that may be
kinematically separable in some analyses.

\subsection{Direct Charged Higgs Production}

As discussed in \cref{ssec:charged}, we have added many charged Higgs search
channels to \HB that are or may --- in the future --- be probed at the LHC\@. The most thoroughly studied (and in
many cases dominant) production channel is the production of a charged Higgs in
association with a top-quark --- denoted $pp \to H^\pm t b$ in the four-flavor
scheme or $bg \to H^\pm t$ in the five-flavor scheme --- which is typically
considered in experimental searches at the LHC, see \eg
Refs.~\cite{Aaboud:2018gjj, Aaboud:2018cwk, Sirunyan:2019hkq, CMS:2019yat}. The
cross section has been calculated including NLO-QCD corrections in the 2HDM and
MSSM~\cite{Berger:2003sm, Dittmaier:2009np, Flechl:2014wfa, Degrande:2015vpa,
	deFlorian:2016spz, Degrande:2016hyf}.

While the 2HDM is the simplest BSM model with a charged Higgs boson $H^\pm$, the
coupling structure of its charged-Higgs--quark couplings readily generalizes
to a large variety of models. In the 2HDM (and also \eg in the MSSM at
tree-level) the relevant charged Higgs coupling to top and bottom quarks has
the form
\begin{equation}
	g_{t\bar{b}H^-}=\sqrt{2}\left(\frac{m_t}{v}P_R \kappa^\pm_t
	+ \frac{m_b}{v}P_L \kappa^\pm_b \right)\label{eq:Hccoup}
\end{equation}
with $\kappa^\pm_t = 1/\tan\beta$, $\kappa^\pm_b = \tan\beta$ for flavor
conserving 2HDM Yukawa sectors of type II (or in the MSSM) and
$\kappa^\pm_t=\kappa^\pm_b=1/\tan\beta$ in Yukawa type I. For a generic
charged Higgs boson $H_j^\pm$ we expand the cross section in terms of
$\kappa^{j\pm}_{t,b}$ defined as in \cref{eq:Hccoup} and obtain
\begin{equation}
	\sigma^{tH_j^-}[m_{H_j^\pm}] = {(\kappa^{j\pm}_t)}^2 \bar{\sigma}^{tH_j^-}_{tt}[m_{H_j^\pm}]
	+ \kappa^{j\pm}_t \kappa^{j\pm}_b \bar{\sigma}^{tH_j^-}_{tb}[m_{H_j^\pm}]
	+ {(\kappa^{j\pm}_b)}^2 \bar{\sigma}^{tH_j^-}_{bb}[m_{H_j^\pm}]\eqdot\label{eq:cxn_tHc_heavy}
\end{equation}
We keep the interference term --- even though it is suppressed by an additional
mass insertion due to the helicity structure of the coupling --- as its contribution
becomes important for charged Higgs masses close to the top-threshold region.

For charged Higgs bosons lighter than the top-quark, the internal top-quark
propagators can go on-shell introducing a dependence on the width of the
top-quark, $\Gamma_t$. According to Ref.~\cite{Degrande:2016hyf}, this can
be approximately included by rescaling the cross section as
\begin{equation}
	\sigma^{tH_j^-}[m_{H_j^\pm}<m_t-m_b] \to {\left(\frac{\Gamma_t^\text{SM}}{\Gamma_t^\text{BSM}}\right)}^2 \sigma^{tH_j^-}[m_{H_j^\pm}<m_t-m_b]\eqdot
\end{equation}
Neglecting decays of the top-quark into first and second generation quarks in
the SM, this ratio of widths is simply given by $\text{BR}(t\to W^+b)$ in the
BSM model under consideration. We therefore parameterize $\sigma^{tH_j^-}$ in
the entire $m_{H_j^\pm}$ range as
\begin{equation}
	\sigma^{tH_j^-}[m_{H_j^\pm}] = \left({(\kappa^{j\pm}_t)}^2 \bar{\sigma}^{tH_j^-}_{tt}[m_{H_j^\pm}]
	+ \kappa^{j\pm}_t \kappa^{j\pm}_b \bar{\sigma}^{tH_j^-}_{tb}[m_{H_j^\pm}]
	+ {(\kappa^{j\pm}_b)}^2 \bar{\sigma}^{tH_j^-}_{bb}[m_{H_j^\pm}]\right)\mathcal{S}\eqcomma\label{eq:cxn_tHc}
\end{equation}
where
\begin{equation}
	\mathcal{S} = \begin{cases}\mathrm{BR}(t\to W^+ b)^2 & \text{if}~\mathrm{BR}(t\to H_j^+ b)>0,\\1&\text{otherwise.}\end{cases}
\end{equation}

We use the results\footnote{\label{twiki}Available at
	\url{https://twiki.cern.ch/twiki/bin/view/LHCPhysics/LHCHXSWGMSSMCharged}.}
of Refs.~\cite{deFlorian:2016spz, Degrande:2016hyf} tabulated in the 2HDM type
II as a function of the charged Higgs mass and $\tan\beta$ and extract the mass
dependent coefficients $\bar{\sigma}_{ab}^{tH_j^-}$ by solving the resulting
system of linear equations. In the region $m_{H_j^\pm}<m_t$ we use
\texttt{HDECAY-6.52}~\cite{Djouadi:1997yw,Djouadi:2018xqq} to calculate the
required branching ratios of the top-quark. The resulting parametrization
reproduces the original results to a relative accuracy of better than $10^{-4}$
for $m_{H_j^\pm}> 2 m_t$ and --- with deviations of at most \SI{2}{\percent} ---
stays well within the theoretical uncertainties of the original
calculation~\cite{Degrande:2016hyf} even for $m_{H_j^\pm}<m_t$.

The parametrization in \cref{eq:cxn_tHc} holds for any charged Higgs boson
$H_j^-$ that has a coupling structure of the form \cref{eq:Hccoup}, and is valid
as long as no other BSM effects contribute up to NLO in QCD\@. In particular
this neglects possible contributions of the form $p p \to b\bar{b} \phi \to
	b\bar{b}H^+ W^-$ that can appear in 2HDM-like models. However, these are
typically only relevant for resonant $\phi$ production where they are treated
separately. SUSY QCD corrections also impact these results, and the dominant
$\Delta_b$ corrections can be included through a rescaling of
$\tan\beta$.\footref{twiki} In the region $m_{H_j^\pm}<m_t$ the approximation
relies on the assumption $\text{BR}(t\to W^+ b) + \text{BR}(t\to H_j^+
	b)\approx1$ (no sum over $j$). If this assumption is violated --- \eg because
the top quark decays into multiple $H^\pm_j$ or into additional new-physics
decay modes --- the threshold behavior would be incorrect and a full
model-specific calculation should be performed. However, heavier $H^\pm_j$ are
insensitive to the top width, and valid cross sections for any number of
$H^\pm_j$ heavier than the top quark can be obtained.

In \HBv{5} this approximation can be accessed through the \Code{HCCS_tHc}
function in the \Code{access_effC.f90} file, which requires $m_{H^\pm}$,
$\kappa^{j\pm}_t$, $\kappa^{j\pm}_b$, and $\text{BR}(t\to H_j^+b)$ as input and
assumes $\text{BR}(t\to W^+b) = 1 - \text{BR}(t\to H_j^+b)$. The function returns
the cross section
\begin{equation}
	\sigma^{tH_j^\pm} = 2\sigma^{tH_j^-},
\end{equation}
as the cross section is charge-symmetric, $\sigma^{tH_j^+} = \sigma^{\bar{t}H_j^-}$. This inclusive value then corresponds to the
required \HB input quantity \Code{CS_Hpmjtb}, see \cref{ssec:charged}.

\subsection{Exclusion Likelihoods for LHC Higgs to
	\boldmath{$\tau^+\tau^-$} Searches}\label{sec:tautau}

In experimental searches for additional Higgs bosons decaying into
$\tau^+\tau^-$ the ATLAS~\cite{Aaboud:2017sjh,Aad:2020zxo} and
CMS~\cite{Sirunyan:2018zut,CMS:2015mca} collaborations have released simplified
exclusion likelihoods as a function of the two contributing single Higgs
production modes, $gg\to \phi$ and $gg\to b\bar{b}\phi$, and for a wide range of
narrow scalar resonance mass hypotheses. The implementation of these nearly
model independent likelihoods in \HB includes an approximate scheme for treating
multiple contributing Higgs bosons of similar mass. The implementation of the
first analysis from LHC Run-1~\cite{CMS:2015mca} for which this input was
provided has been described in detail in Ref.~\cite{Bechtle:2015pma}. We present
the implementation and validation of new LHC Run-2 analyses and discuss
improvements to the derivation of exclusion limits from the provided likelihood
information. More details on the underlying likelihood reconstruction method can
be found in Ref.~\cite{Bechtle:2015pma}.

The profiled likelihood analyses underlying the experimental results use the
test statistic
\begin{equation}
	q_\mu = -2\ln\frac{\mathcal{L}(N|\mu\cdot s(m) + b, \hat{\theta}_\mu)}{\mathcal{L}(N|\hat{\mu}\cdot s(m) + b, \hat{\theta})}\eqcomma
\end{equation}
where $N$ is the observed data, $b$ is the background expectation, and $s(m)$ is
the signal expectation for a given hypothesized resonance mass $m$ and given
contributions of the two sub-channels. A limit is set on the signal strength
modifier $\mu$ in the presence of the globally optimized nuisance parameters
$\hat{\theta}$, the globally optimized signal strength $\hat{\mu}$
and the conditionally optimized nuisance parameters
$\hat{\theta}_{\mu}$ for the given value of $\mu$. The experiments
provide expected, $q_\mu^\text{exp}$, and observed, $q_\mu^\text{obs}$, values
for this test as a function of the two sub-channel contributions for different
resonance masses.

Since the likelihood was parametrized in terms of a single narrow scalar
resonance, in a specific model application in \HB multiple Higgs bosons
potentially contributing to the signal have to be combined and mapped onto the
likelihood parametrization. This is done in \HB by a \emph{clustering}
algorithm. For this all Higgs bosons within
\begin{equation}
	|m_i - m_j| \leq \Delta_\text{res} \cdot  \max(m_i,m_j)
\end{equation}
are combined into a cluster, their rates are incoherently
summed\footnote{Interference effects of Higgs bosons can be accounted for by
	providing channel rates (see \cref{ssec:channelrates}) as theoretical input.},
and a signal-rate-weighted cluster mass is used to approximate the mass of the
single resonance mass. The numerical coefficient $\Delta_\text{res}$ is chosen
to approximately match the mass resolution of the $\tau^+\tau^-$ channel under
consideration, and is currently set to $20\%$ for all implemented analyses. This
algorithm has already successfully been applied in various analyses, including
cases in which more than two Higgs bosons form a
cluster~\cite{Bechtle:2016kui,Bahl:2018zmf,Bechtle:2015pma}.

In the limit of large numbers the test statistic $q_\mu$ can be treated as a
$\Delta\chi^2$ above minimum such that \CL{68} and \CL{95} exclusion bounds are
obtained at $q_\mu=2.28$ and $5.99$, respectively, corresponding to a two-sided
limit (or fit). This approach was employed by the experimental collaborations to
obtain the confidence regions in the presented two-dimensional cross section
planes for fixed resonance mass, and hence also used to obtain \CL{95} exclusion
limits from the likelihood information in \HBv{4}~\cite{Bechtle:2015pma}. More
appropriate for limit setting, however, is a one-sided \emph{upper} limit on the
signal cross section. Therefore, \HBv{5} uses an improved approach in which \CLs
is directly calculated from the provided likelihood information and the \CL{95}
allowed region is obtained at $\CLs>0.05=1-\SI{95}{\percent}$. This
reconstruction relies on the fact that the quantity $q_\mu^\text{exp}$ provided
by the experiments can be interpreted as\footnote{We are very grateful to Artur
	Gottmann and Roger Wolf for suggesting this approach to us.}
\begin{equation}
	q_\mu^\text{exp}=\frac{\mu^2}{\sigma^2}\eqcomma
\end{equation}
where $\sigma$ is the expected effective Gaussian uncertainty of the signal strength
modifier $\mu$. The expected and observed \CLs can
then be obtained just from $q_\mu^\text{exp}$ and $q_\mu^\text{obs}$ in the
asymptotic limit as
\begin{equation}
	\CLs = \frac{\text{CL}_{s+b}}{\text{CL}_b}\label{eq:CLs}
\end{equation}
using~\cite{Cowan:2010js}
\begin{align}
	\text{CL}^\text{exp}_{s+b} & = 1 - \Phi\left(\sqrt{q_\mu^\text{exp}}\right)\eqcomma \\
	\text{CL}^\text{exp}_b     & = 0.5\eqcomma
\end{align}
for the expected limit and
\begin{align}
	\text{CL}^\text{obs}_{s+b} & = 1 - \begin{cases}
		\mathrlap{\Phi\left({\scriptstyle\sqrt{q_\mu^\text{exp}}}\right)}\hphantom{\Phi\left({\scriptstyle\sqrt{q_\mu^\text{obs}}-\sqrt{q_\mu^\text{exp}}}\right)} & 0 < q^\text{obs}_\mu \leq q^\text{exp}_\mu\eqcomma \\
		\Phi\left(\frac{q_\mu^\text{obs}+q_\mu^\text{exp}}{2\sqrt{q_\mu^\text{exp}}}\right)                                                                        & q^\text{obs}_\mu > q^\text{exp}_\mu\eqcomma
	\end{cases} \\
	\text{CL}^\text{obs}_b     & = 1 - \begin{cases}
		\Phi\left({\scriptstyle\sqrt{q_\mu^\text{obs}}-\sqrt{q_\mu^\text{exp}}}\right)       & 0 < q^\text{obs}_\mu \leq q^\text{exp}_\mu\eqcomma \\
		\Phi\left(\frac{q_\mu^\text{obs}-q_\mu^\text{exp}}{2\sqrt{q_\mu^\text{exp}}} \right) & q^\text{obs}_\mu > q^\text{exp}_\mu\eqcomma        \\
	\end{cases}
\end{align}
for the observed limit. In all of these, $\Phi$ denotes the cumulative normal
distribution function. The experimental collaborations use the $\CLs>0.05$
criterion for model-specific limit setting, \eg in the context of MSSM benchmark
scenarios. \HBv{5} uses the improved approach that directly employs $\CLs>0.05$
instead of $\Delta\chi^2<5.99$ both for determining the sensitivity of the
analysis via $\CLs^\text{exp}$ and for obtaining the \CL{95} limit via
$\CLs^\text{obs}$. This improved methodology in \HB that more closely resembles
the one employed by the experimental collaborations enables a reliable
reconstruction of limits from the provided likelihood information. As we will
demonstrate below, with \HB it is now possible to even reproduce and understand
methodical differences between the official ATLAS and CMS model interpretations.

\begin{figure}[t]
	\includegraphics[width = \textwidth]{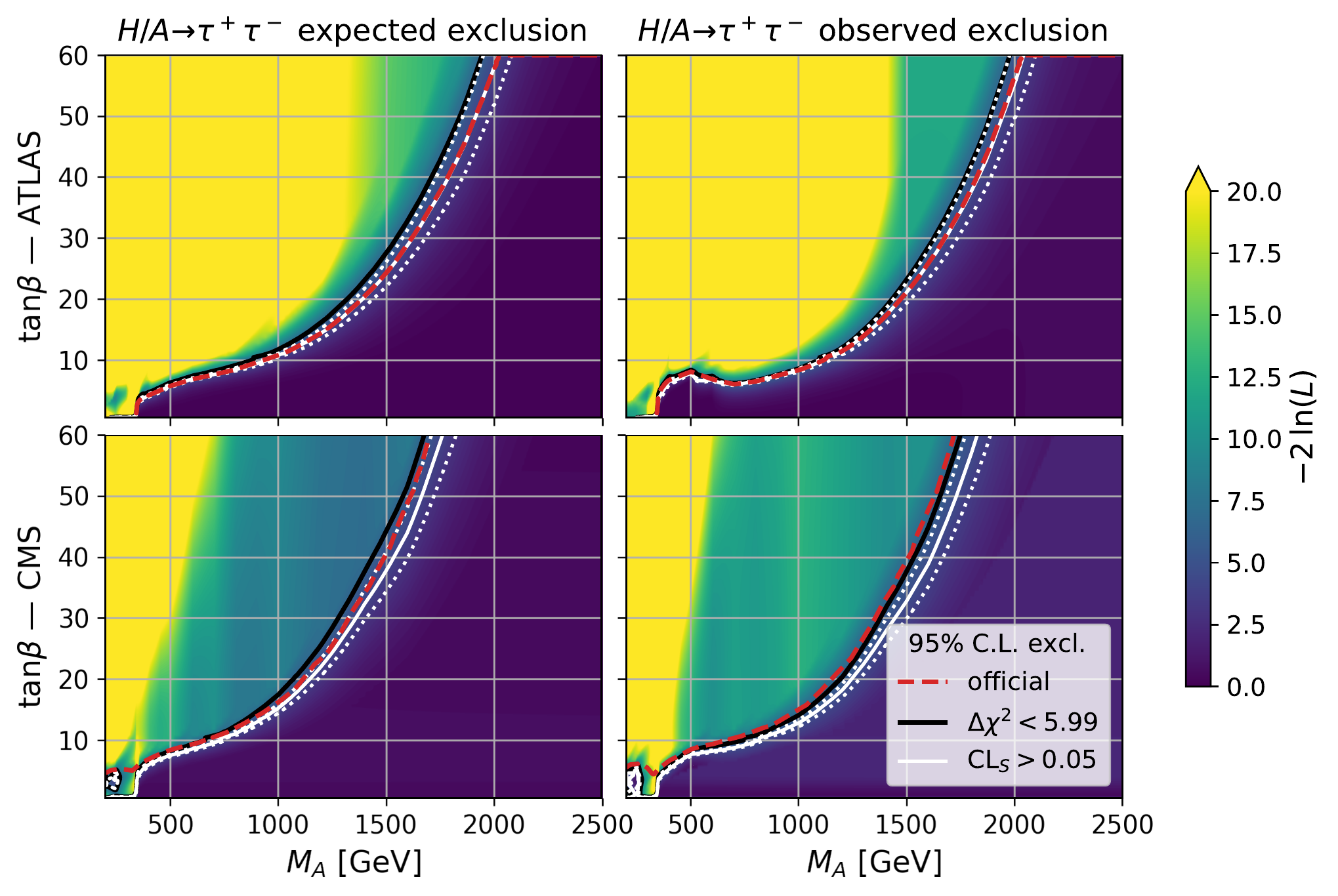}
	\caption{Expected (\emph{left panels}) and observed (\emph{right panels})
	exclusion likelihood from LHC $pp\to H/A \to \tau^+\tau^-$ searches at
	ATLAS~\cite{Aad:2020zxo} (\emph{top panels}) and CMS~\cite{Sirunyan:2018zut}
	(\emph{bottom panels}) in the $M^h_{125}$ scenario~\cite{Bahl:2018zmf}. The
	solid black and white lines show the reconstructed \CL{95} limit in \HB
	using the old (black, based on $\Delta\chi^2<5.99$) and the improved new
	method (white, based on $\CLs>0.05$). For the limit based on $\CLs>0.05$ the
	white-dotted lines indicate the variation of the limit due to
	signal-model-dependent theory uncertainties. The red dashed line shows the
	official ATLAS or CMS \CL{95} limit.}
	\label{fig:exclLLH}
\end{figure}

\Cref{fig:exclLLH} shows a validation of the \HB likelihood reconstruction
algorithm against the most recent \SI{13}{\TeV} experimental analyses by
ATLAS~\cite{Aad:2020zxo} (at \SI{139}{\per\fb}, top) and
CMS~\cite{Sirunyan:2018zut} (at \SI{36}{\per\fb}, bottom) in the $M^h_{125}$
scenario~\cite{Bahl:2018zmf} of the MSSM\@. The theoretical predictions for the
scenario are taken from the LHCHXSWG, based on the following
prescription~\cite{Heinemeyer:2013tqa, deFlorian:2016spz}:
\texttt{FeynHiggs}~\cite{Heinemeyer:1998yj, Heinemeyer:1998np, Degrassi:2002fi,
	Frank:2006yh, Hahn:2013ria, Bahl:2016brp, Bahl:2017aev} is taken for the
calculation of the MSSM masses and couplings, a combination of
\texttt{FeynHiggs} and \texttt{HDECAY}~\cite{Djouadi:1997yw, Djouadi:2018xqq}
for the $\tau^+\tau^-$ branching ratio. The gluon fusion cross section
predictions are obtained with \texttt{SusHi}~\cite{Harlander:2012pb,
	Harlander:2016hcx} including all available higher order
corrections~\cite{Spira:1995rr, Harlander:2002wh, Harlander:2002vv,
	Anastasiou:2002wq, Anastasiou:2002yz, Ravindran:2003um, Harlander:2004tp,
	Harlander:2005rq, Degrassi:2010eu, Degrassi:2011vq, Degrassi:2012vt}, and the
$b\bar{b}$ associated channel uses matched cross section
predictions~\cite{Harlander:2003ai, Dittmaier:2003ej, Dawson:2003kb,
	Bonvini:2015pxa, Bonvini:2016fgf, Forte:2015hba, Forte:2016sja}. The
prescription for the theoretical predictions in the MSSM benchmark scenarios is
such that the model-dependent theoretical uncertainties of the predicted masses,
cross sections and branching ratios should be incorporated by the user through
appropriate variations of the theoretical predictions. We will discuss the
impact of these theoretical uncertainties on the obtained limits below.

The color code in \cref{fig:exclLLH} shows the expected (observed) reconstructed
likelihood value $-2\ln(L)$ from \HB\ on the left (right). The black contour
indicates the \HB \CL{95} exclusion limit reconstructed using the old
$\Delta\chi^2<5.99$ approach, while the solid white contour displays the \CL{95}
limit from the new \CLs method. The reconstructed limits displayed by the
solid black and white contours do not take into account
signal-model-dependent theoretical uncertainties. For the limit based on
the \CLs method the white-dotted lines indicate the uncertainty band around the
solid white contour according to the model-dependent theoretical
uncertainties in the considered
$M^h_{125}$ benchmark scenario of the MSSM~\cite{Bahl:2018zmf}, see below. For
comparison, the red-dashed lines show the official \CL{95} limits from ATLAS
(upper panels) and CMS (lower panels).

We start by comparing the limits obtained with the $\Delta\chi^2$ and the \CLs
method (solid black and white contours). The $\CLs>0.05$ criterion results in a
larger excluded area compared to the $\Delta\chi^2<5.99$ criterion. This feature
is expected and can be understood as follows. The $\Delta\chi^2=5.99$ limit
corresponds to a \CL{95} limit on $\text{CL}_{s+b}$ in the Gaussian
approximation. However, in order to prevent erroneous exclusions in regions
where the search has no sensitivity~\cite{Read:2002hq} \CLs is constructed by
dividing $\text{CL}_{s+b}$ by $\text{CL}_b$, see \cref{eq:CLs}. Since the
expectation value of $\text{CL}_b$ in the absence of any signal is $\langle
\text{CL}_b\rangle=\num{0.5}$, a \CL{95} limit on \CLs approximately corresponds
to a $1-\num{0.05}/\langle\text{CL}_b\rangle\approx\CL{90}$ limit on
$\text{CL}_{s+b}$.\footnote{In the Gaussian approximation and for the
two-dimensional case considered here this would correspond to a
$\Delta\chi^2<4.61$ limit.}

We now compare the limit obtained with the \CLs method with the results obtained
by ATLAS and CMS in their analyses for the $M^h_{125}$ benchmark scenario. As
explained above, the reconstructed limit in \HB is based on the (nearly)
model-independent likelihood provided by ATLAS and CMS which by construction
does not contain any model-specific theoretical uncertainties on the cross
sections and branching ratios of the signal processes. Accordingly, the solid
white contour corresponds to the limit obtained with the \CLs method without
taking into account model-specific theoretical uncertainties. We find that the
resulting limit agrees almost perfectly with the one obtained in the ATLAS
analysis, both for the expected (left upper panel) and the observed limit (right
upper panel). On the other hand, the benchmark analysis of CMS excludes a
smaller region than in our \CLs analysis (solid white contour), both for the
expected and the observed limit (lower panels). As we have verified via direct
communication with members of the ATLAS and CMS
collaborations~\cite{privateCommunicationBilletal}, this feature can be
understood from the fact that the experimental interpretation of the benchmark
scenario by CMS includes model-specific theoretical uncertainties on the signal
cross-sections, while in the ATLAS analysis no such signal-model-dependent
theoretical uncertainties have been taken into account.

As a final step of this comparison we now take into account
signal-model-dependent theoretical uncertainties with \HB. The dotted white
contours indicate the uncertainty band around the solid white contour that has
been obtained by running \HB with input rates at the upper and lower end of the
theoretical uncertainties and interpreting the resulting difference in the
\CL{95} exclusion as a theoretical error band. In line with the CMS analysis, we
include scale and parton distribution function uncertainties on the cross
section predictions, provided by the LHCHXSWG for the $M^h_{125}$ scenario.
Theoretical uncertainties on the BRs are not considered since they are
negligible by comparison~\cite{deFlorian:2016spz}. The upper branch of this band
shows how the limit is weakened by the incorporation of the
signal-model-dependent theoretical uncertainties. We find that this contour
agrees well with the limit that has been obtained in the CMS benchmark analysis.
We expect that an ATLAS limit incorporating the signal-model-dependent
uncertanties would be less constraining and closer to the upper white-dotted
limit obtained with \HB.\footnote{In \cref{fig:exclLLH} the black contour
indicating the \HB limit using the $\Delta\chi^2$ method in all cases happens to
be close to the limit that incorporates the signal-model-dependent uncertainties
(and thus close to the official CMS results). We stress that this is a
scenario-dependent coincidence and that the theoretical uncertainties are not
captured by the $\Delta\chi^2$ limit.}

By default, the likelihood information is used to reconstruct a \CL{95} limit
that is then treated like any other limit in \HB. \HB selects the \emph{Higgs
	cluster} giving the largest expected exclusion likelihood as the \emph{most
	sensitive Higgs boson combination} to be tested against the observation. The
advantage of reconstructing the limit from the likelihood is that the full
efficiency information on the two involved production channels is incorporated.
This is especially important in the BSM $\phi\to\tau^+\tau^-$ channel, since the
relative contributions of $gg\to \phi$ and $gg\to b\bar{b}\phi$ can change
drastically through the model parameter space, \eg in the MSSM or the 2HDM\@.
The value of the likelihood can also be accessed directly through the
\Code{HiggsBounds_get_likelihood} subroutines (see online documentation), such
that it can be included in a global likelihood analysis of BSM models, see \eg
Refs.~\cite{Bechtle:2016kui, Bagnaschi:2017tru, Costa:2017gup,
	Bagnaschi:2018zwg}.

\begin{figure}
	\centering
	\includegraphics[width=0.9\textwidth]{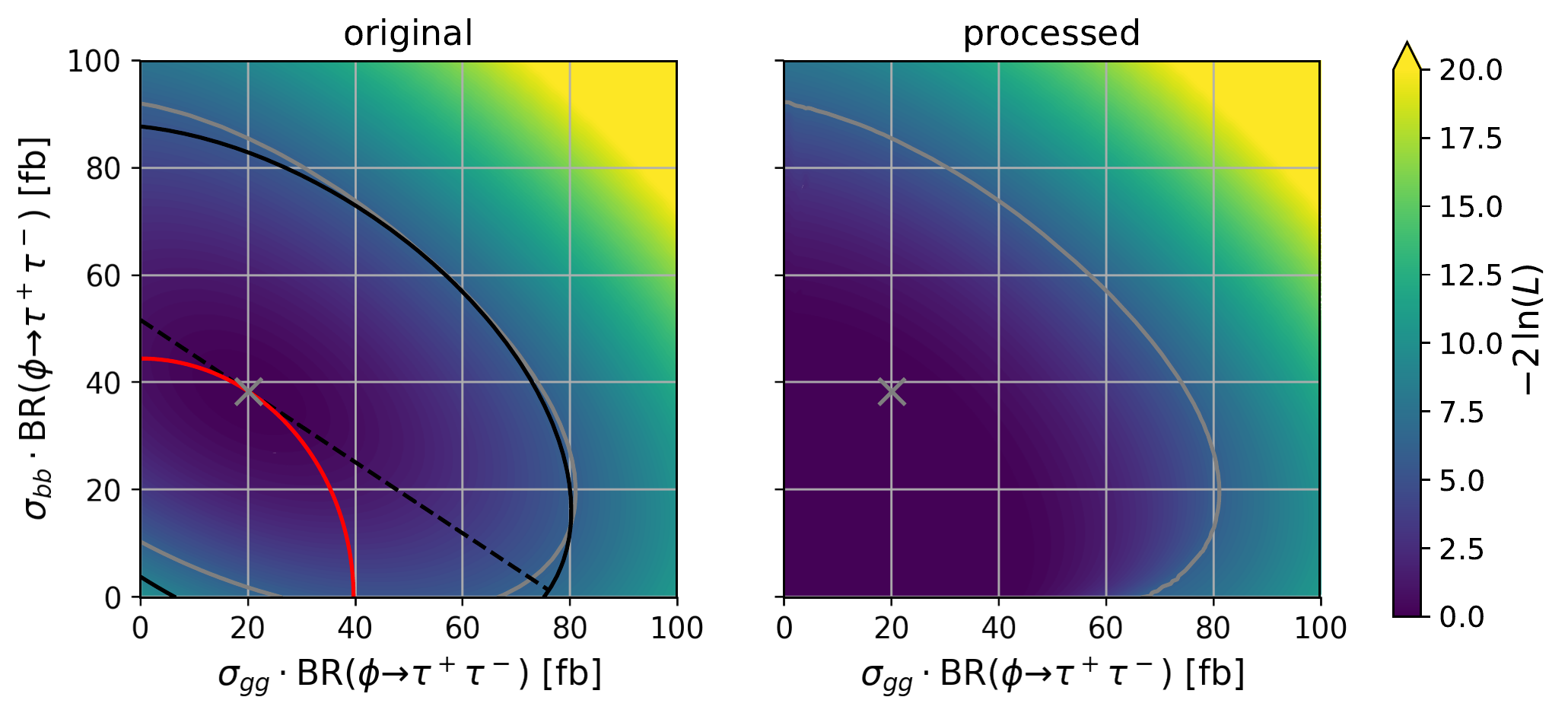}
	\caption{Likelihood of the ATLAS search~\cite{Aad:2020zxo} in the plane of
		$\sigma_{gg}\cdot \mathrm{BR}(\phi\to\tau^+\tau^-)$ and
		$\sigma_{bb}\cdot \mathrm{BR}(\phi\to\tau^+\tau^-)$ for
		$M_\phi=\SI{400}{\GeV}$. The color code in the left panel shows the
		likelihood as provided by ATLAS, while the right panel shows the
		processed likelihood used in \HB. The \CL{95} contour and best fit point
		are indicated in grey. The black ellipse with the corresponding
		long axis (dashed) and the red ellipse shown in the left panel are used
		in the construction of the processed likelihood (see text).}
	\label{fig:atlas_fudge}
\end{figure}

The implementation of the recent ATLAS result~\cite{Aad:2020zxo} raised a
different kind of issue, as it contains an observed excess with a local
significance of more than $2\sigma$ for masses in the range
\SIrange{400}{500}{\GeV}. Therefore, the parameter point of zero signal rates
lies outside the \CL{95} contour of the provided likelihoods. A direct
application of these results in the original form to a model would exclude
parameter points that feature scalar boson(s) with small or vanishing $pp\to
	\phi \to \tau^+\tau^-$ rates in that mass range. At the same time, parameter
points with no scalars in that mass range would not be excluded, as the
likelihood would not be evaluated if no scalar boson is within the mass range.
We therefore use an approximate approach to avoid this inconsistent behavior
while keeping the overall features of the likelihood profile intact. This
approach is applied to the mass range with an excess of more than $1\sigma$, \ie
for the mass planes at \SIlist{350;400;500}{\GeV} provided by ATLAS\@. It
restores the property $q_\mu^\text{obs}=0$ for $\hat{\mu} > \mu$ required of a
test statistic used in limit setting~\cite{Cowan:2010js}, where $\hat{\mu}$ is the best fit rate.

We use the $M_\phi=\SI{400}{\GeV}$ mass plane, shown in \cref{fig:atlas_fudge},
to illustrate the approach. The grey contours are the \CL{95},
$\Delta\chi^2=5.99$ contours\footnote{We use the simpler $\Delta\chi^2$ approach
only to construct the processed likelihood tables. The \CLs method described
above is always used to obtain \CL{95} limits in \HBv{5}.} of the original
(processed, \ie based on our approach for avoiding the exclusion of \emph{too
small} BSM rates) observed likelihood profile $q_\mu^\text{obs}$ shown on the
left (right). We first fit an ellipse centered at the best fit point (grey
cross) to the \CL{95} contour of the original $q_\mu^\text{obs}$. In
\cref{fig:atlas_fudge} (left) this ellipse is shown in black. We then construct
the ellipse shown in red, which is centered at the origin with axes parallel to
the coordinate axes. We fix the eccentricity by requiring this ellipse to be
tangential to the long axis of the black ellipse in the best fit point (the
black-dashed line). We consider all signal rates on this red ellipse to be equal
to the best fit rate, $\mu\sim\hat\mu$.\footnote{To first approximation, the
ratio between the long and short axes of this ellipse resembles the ratio of
signal efficiencies in the two production channels. Therefore, our construction
indeed approximately determines the parameter region where the total fiducial
signal rate is equal to the one at the best fit point.} Accordingly, the
likelihood inside the red ellipse is set to zero, since lower predicted rates
than at the best-fit point should not be disfavored in a limit setting
procedure. To obtain a smooth transition, we introduce polar coordinates
$(r,\theta)$ and, for each angle $\theta$, approximate the likelihood profile by
\begin{equation}
	-2\ln(L) = q^\text{obs}_\mu \to \begin{cases}
		0                                                            & \text{for }r\leq r_E \Leftrightarrow \mu \leq \hat\mu \eqcomma \\
		5.99 \left(\frac{r - r_\text{E}}{r_{95}-r_\text{E}}\right)^2 & \text{for }r_E < r < r_{95} \eqcomma                           \\
		q^\text{obs}_\mu                                             & \text{for }r_{95}\leq r \eqcomma
	\end{cases}
\end{equation}
where $r_\text{E}$ ($r_{95}$) is the radial component of the red ellipse (the
grey \CL{95}, $\Delta\chi^2=5.99$ contour) for the given $\theta$. Outside the
grey contour the likelihood remains unchanged from the original. This leads to
the processed likelihood profile shown in the right panel of
\cref{fig:atlas_fudge}. The upper part of the grey \CL{95} contour and the
likelihood for larger rates remain unchanged, the inconsistent exclusion of low
and vanishing rates is avoided, and the intermediate region is continuously
interpolated. These processed likelihood profiles are used instead of the
original ones in \HB. If \HB is used in a global fit and only the likelihood
values, but no corresponding reconstructed limits are desired, the original
values can be used instead by setting the logical \Code{preventOverexclusion}
parameter in \Code{likelihoods.F90} (see online documentation) to false.

\section{User Operating Instructions}
\label{sec:user}

For \HBv{5} we have made substantial changes on the structure and online
platform of the \HB source code. The code has moved to a \textsc{GitLab}
repository and is now available at
\url{https://gitlab.com/higgsbounds/higgsbounds}. This modernization effort also
included moving to CMake as the build system. \HB is now compiled by running:
\begin{verbatim}
mkdir build && cd build
cmake ..
make
\end{verbatim}
The only requirements are CMake and a Fortran compiler. This will compile the
library, the main executable and a number of example programs that illustrate
different use cases. More detailed information on building and linking \HB can
be found on the above-mentioned website.

\subsection{\Code{Fortran} Subroutines}
\label{ssec:subroutines}

The \texttt{Fortran} subroutines are the most powerful and versatile way of
using \HB. Up-to-date descriptions of the various \texttt{Fortran} subroutines
and functions can be found in the online documentation at
\url{https://higgsbounds.gitlab.io/higgsbounds}.

Compared to \HBv{4}, all of the input subroutines have been extended to include
all of the quantities discussed in \cref{sec:theo}. The input subroutine
arguments are named as in
\cref{tab:hadrneutralHiggsinput,tab:lepneutralxs,tab:brneutral,tab:effCneutral,tab:chargedHiggsinput}
and are either \texttt{double precision} or \texttt{integer} arrays with
dimensions given by the number of neutral Higgs bosons and/or the number of
charged Higgs bosons.

As a further improvement, \HBv{5} includes a \texttt{C} interface to all of the
\texttt{Fortran} subroutines to facilitate the use of \HB from \texttt{C} or
\texttt{C++} codes. This interface automatically handles type conversion and
accounts for the different storage orders of multidimensional arrays between
\texttt{C} and \texttt{Fortran}. The C interface is included in the online
documentation.

\subsection{Command-line Version}
\label{ssec:commandline}
Compiling \HB generates a main executable that can be run as
\begin{verbatim}
./HiggsBounds <whichanalyses> <whichinput> <nHzero> <nHplus> <prefix>
\end{verbatim}
where the arguments specify the following: \Code{<whichanalyses>} specifies
which experimental data is selected for the model test --- `\Code{LandH}' for
all implemented results, `\Code{onlyL}' for LEP results only, `\Code{onlyH}' for
hadron-collider results only, and `\Code{onlyP}' for published results only;
\Code{<whichinput>} specifies whether the model input on the production and
decay rates is provided in the effective couplings approximation
(`\Code{effC}'), at the cross section level (`\Code{hadr}'), or via an SLHA
input file (`\Code{SLHA}'). The arguments \Code{<nHzero>} and \Code{<nHplus>}
specify the number of neutral and charged Higgs bosons, respectively. The
argument \Code{<prefix>} denotes the path to the input files including any part
of the filename that is common to all input files.

Additionally an executable called \Code{AllAnalyses} is generated. It prints a
table listing all of the experimental analyses implemented in \HB, including
\textsf{arXiv} identifier or report numbers as well as \textsf{InspireHEP}
cite keys. Since we continuously implement new experimental results in the
code, we refer to this executable for an up-to-date list of what is included in
the version of \HB that is being used. A bibliography file that includes
entries for all implemented analyses is available on the website.

\subsubsection{\HB Data Files Input}

\begin{table}[p]
	\centering
	\renewcommand{\arraystretch}{1.05}
	\footnotesize
	\begin{tabularx}{\textwidth}{lccP{0mm}l}
		\toprule
		Data file name                                                     & \makebox[3mm][c]{\Code{effC}} & \makebox[3mm][c]{\Code{hadr}} & \multicolumn{2}{l}{Contents}                                                                        \\
		\midrule
		\Code{MH_GammaTot.dat}                                             & y                             & y                             & k,                           & \Code{Mh, MhGammaTot}                                                \\[1mm]
		\Code{MHplus_GammaTot.dat}                                         & y                             & y                             & k,                           & \Code{Mhplus, MhplusGammaTot}                                        \\[1mm]
		\Code{MHall_uncertainties.dat}                                     & o                             & o                             & k,                           & \Code{dMh, dMhplus}                                                  \\[1mm]
		\Code{CP_values.dat}                                               & n                             & y                             & k,                           & \Code{CP_value}                                                      \\[1mm]
		\Code{effC.dat}                                                    & y                             & n                             & k,                           & \Code{ghjss_s, ghjss_p, ghjcc_s, ghjcc_p,}                           \\
		                                                                   &                               &                               &                              & \Code{ghjbb_s, ghjbb_p, ghjtt_s, ghjtt_p,}                           \\
		                                                                   &                               &                               &                              & \Code{ghjmumu_s, ghjmumu_p, ghjtautau_s, ghjtautau_p,}               \\
		                                                                   &                               &                               &                              & \Code{ghjWW, ghjZZ, ghjZga, ghjgaga, ghjgg, ghjhiZ}$^*$              \\
		\Code{BR_H_OP.dat}                                                 & o                             & y                             & k,                           & \Code{BR_hjss, BR_hjcc, BR_hjbb, BR_hjtt,}                           \\
		                                                                   &                               &                               &                              & \Code{BR_hjmumu, BR_hjtautau, BR_hjWW,}                              \\
		                                                                   &                               &                               &                              & \Code{BR_hjZZ, BR_hjZga, BR_hjgaga, BR_hjgg}                         \\[1mm]
		\Code{BR_H_NP.dat}                                                 & y                             & y                             & k,                           & \Code{BR_hjinvisible, BR_hkhjhi}$^*$\Code{, BR_hjhiZ}$^*$\Code{,}    \\
		                                                                   &                               &                               &                              & \Code{BR_hjemu, BR_hjetau, BR_hjmutau, BR_hjHpiW}                    \\[1mm]
		\Code{BR_t.dat}                                                    & y                             & y                             & k,                           & \Code{BR_tWpb, BR_tHpb}                                              \\[1mm]
		\Code{BR_Hplus.dat}                                                & y                             & y                             & k,                           & \Code{BR_Hpcs, BR_Hpcb, BR_Hptaunu, BR_Hptb,}                        \\
		                                                                   &                               &                               &                              & \Code{BR_HpWZ, BR_HpjhiW}                                            \\[1mm]
		\Code{additional.dat}                                              & o                             & o                             & k,                           & \dots                                                                \\
		\Code{LEP_HZ_CS_ratios.dat}                                        & y                             & y                             & k,                           & \Code{CS_lep_hjZ_ratio}                                              \\[1mm]
		\Code{LEP_H_ff_CS_ratios.dat}                                      & y                             & y                             & k,                           & \Code{CS_lep_bbhj_ratio, CS_lep_tautauhj_ratio}                      \\[1mm]
		\Code{LEP_2H_CS_ratios.dat}                                        & y                             & y                             & k,                           & \Code{CS_lep_hjhi_ratio}$^*$                                         \\
		\Code{LEP_HpHm_CS_ratios.dat}                                      & y                             & y                             & k,                           & \Code{CS_lep_HpjHmj_ratio}                                           \\
		\emph{coll}\Code{_1H_hadCS_ratios.dat}                             & n                             & y                             & k,                           & \Code{CS_hj_ratio, CS_gg_hj_ratio, CS_bb_hj_ratio,}                  \\
		(\emph{coll} = \Code{TEV}, \Code{LHC7}, \Code{LHC8}, \Code{LHC13}) &                               &                               &                              & \Code{CS_hjW_ratio, CS_hjZ_ratio, CS_vbf_ratio,}                     \\
		                                                                   &                               &                               &                              & \Code{CS_tthj_ratio,} \Code{CS_thj_tchan_ratio, CS_thj_schan_ratio}  \\
		                                                                   &                               &                               &                              & \Code{CS_tWhj_ratio,} \Code{CS_qq_hjZ_ratio,} \Code{CS_gg_hjZ_ratio} \\[1mm]
		\emph{coll}\Code{_Hplus_hadCS.dat}                                 & y                             & y                             & k,                           & \Code{CS_Hpjtb, CS_Hpjcb, CS_Hpjbjet, CS_Hpjcjet,}                   \\
		(\emph{coll} = \Code{LHC8}, \Code{LHC13})                          &                               &                               &                              & \Code{CS_Hpjjetjet, CS_HpjW, CS_HpjZ, CS_vbf_Hpj,}                   \\
		                                                                   &                               &                               &                              & \Code{CS_HpjHmj, CS_Hpjhi}                                           \\[1mm]
		\bottomrule
	\end{tabularx}
	\caption{File names and data format for the contents of \HB input files. The
		right column shows the order of the input data arrays within one row of the
		input file (\Code{k} is the line number). See the text for details on the
		order of elements within the arrays and the handling of symmetric
		multidimensional arrays (marked with $^*$). The middle columns indicate
		whether the files are required in the effective couplings approximation
		(\Code{effC}) or hadronic cross section (\Code{hadr}) input scheme [y(es),
				n(o), o(ptional)].}
	\label{table:contentsoffiles1}
\end{table}

If \HB is run from the command line with the option \Code{<whichinput>=effC} or
\Code{hadr} the model input needs to be specified via \HB specific input files.
These are white\-space separated tabular text files containing the input
quantities for one datapoint per row. With respect to \HBv{4}, the input files
have been adjusted to the changes in the input quantities detailed in
\cref{sec:theo}.

An overview of all data input files and their data structure is given in
\cref{table:contentsoffiles1}. For some higher-dimensional arrays only some
elements have to be specified, as will be explained below.
\Cref{table:contentsoffiles1} also specifies whether the data file is required
for either of the two \HB input schemes (\Code{effC} or \Code{hadr}), or used as
optional input. If a required file is not provided as input, \HB warns the user
but proceeds to run while setting the unspecified input quantities to zero. In
\cref{table:contentsoffiles1} we assume that both neutral and charged Higgs
bosons are present in the model. Obviously, if either the number of neutral or
charged Higgs bosons is zero, the corresponding input files are also not
required.

For the two-dimensional input arrays \Code{ghjhiZ} and \Code{CS_lep_hjhi} only
the lower left triangle (including the diagonal) is required, since they are
symmetric matrices. As an example, for three neutral Higgs bosons ($N_{h^0} =
	3$) the symmetric matrix \Code{A},
\begin{equation}
	\Code{A} = \begin{pmatrix} \Code{A[1,1]} & \gray{\Code{A[1,2]}} & \gray{\Code{A[1,3]}} \\
		\Code{A[2,1]} & \Code{A[2,2]}        & \gray{\Code{A[2,3]}} \\
		\Code{A[3,1]} & \Code{A[3,2]}        & \Code{A[3,3]}\end{pmatrix}\eqcomma
\end{equation}
should be specified in the input file in the order
\begin{equation*}
	\Code{A[1,1]},\ \Code{A[2,1]},\ \Code{A[2,2]},\ \Code{A[3,1]},\ \Code{A[3,2]},\ \Code{A[3,3]}\eqdot
\end{equation*}
In contrast, for the two-dimensional input array \Code{BR_hjhiZ}, all off-diagonal elements need to be specified. Again for the $N_{h^0} = 3$ example, we have
\begin{equation}
	\Code{BR_hjhiZ} = \begin{pmatrix} \gray{\Code{BR_hjhiZ[1,1]}} & \Code{BR_hjhiZ[1,2]}        & \Code{BR_hjhiZ[1,3]}        \\
		\Code{BR_hjhiZ[2,1]}        & \gray{\Code{BR_hjhiZ[2,2]}} & \Code{BR_hjhiZ[2,3]}        \\
		\Code{BR_hjhiZ[3,1]}        & \Code{BR_hjhiZ[3,2]}        & \gray{\Code{BR_hjhiZ[3,3]}}\end{pmatrix}\eqcomma
\end{equation}
thus, the elements should be specified as
\begin{align*}
	 & \Code{BR_hjhiZ[1,2]},\ \Code{BR_hjhiZ[1,3]},\ \Code{BR_hjhiZ[2,1]},\ \Code{BR_hjhiZ[2,3]}, \\
	 & \Code{BR_hjhiZ[3,1]},\ \Code{BR_hjhiZ[3,2]}\eqdot
\end{align*}
The three-dimensional input array \Code{BR_hkhjhi[k,j,i]} is symmetric under
exchange of the final state Higgs boson indices \Code{i} and \Code{j} and
elements with \Code{k=j} or \Code{k=i} are zero ($k$ is the index of the
decaying Higgs boson). The $N_{h^0}^2 (N_{h^0}-1)/2$ non-redundant elements can
be specified in the following way: For every $\Code{k}\in \{1,N_{h^0}\}$ we
specify the lower left triangle (including the diagonal), but with the
\Code{k}th column and  \Code{k}th row removed, \eg for $N_{h^0} = 3$
\begin{align*}
	\Code{BR_hkhjhi[1,2,2]},~\Code{BR_hkhjhi[1,3,2]},~\Code{BR_hkhjhi[1,3,3]}, \\
	\Code{BR_hkhjhi[2,1,1]},~\Code{BR_hkhjhi[2,3,1]},~\Code{BR_hkhjhi[2,3,3]}, \\
	\Code{BR_hkhjhi[3,1,1]},~\Code{BR_hkhjhi[3,2,1]},~\Code{BR_hkhjhi[3,2,2]}.
\end{align*}
The input arrays \Code{BR_hjHpiW}, \Code{BR_HpjhiW} and \Code{CS_Hpjhi} are not
reducible and should be specified row by row in the input files.

\subsubsection{SLHA}

\begin{table}[t]
	\centering
	\small
	\texttt{
		\begin{tabular}{rrrrrl}
			\toprule
			\multicolumn{4}{l}{Block ChargedHiggsLHC13} & \texttt{\#} & (in pb)                                                             \\
			5                                           & 6           & 37      & 1.2800 & \texttt{\#} & t-b-Hpm production                 \\
			4                                           & 5           & 37      & 0.4180 & \texttt{\#} & c-b-Hpm production                 \\
			2                                           & 5           & 37      & 0.0002 & \texttt{\#} & u-b-Hpm production                 \\
			3                                           & 4           & 37      & 0.5100 & \texttt{\#} & c-s-Hpm production                 \\
			1                                           & 4           & 37      & 1.1200 & \texttt{\#} & c-d-Hpm production                 \\
			1                                           & 2           & 37      & 0.0001 & \texttt{\#} & u-d-Hpm production                 \\
			2                                           & 3           & 37      & 0.0010 & \texttt{\#} & u-s-Hpm production                 \\
			0                                           & 24          & 37      & 0.0150 & \texttt{\#} & W-Hpm production                   \\
			0                                           & 23          & 37      & 0.0150 & \texttt{\#} & Z-Hpm production                   \\
			1                                           & 1           & 37      & 0.0000 & \texttt{\#} & Hpm vector-boson-fusion production \\
			0                                           & -37         & 37      & 0.0003 & \texttt{\#} & HpHm production                    \\
			0                                           & 25          & 37      & 0.0005 & \texttt{\#} & Hpmh0 production                   \\
			0                                           & 35          & 37      & 0.0002 & \texttt{\#} & HpmH0 production                   \\
			0                                           & 36          & 37      & 0.0004 & \texttt{\#} & HpmA0 production                   \\
			\bottomrule
		\end{tabular}}
	\caption{Example for the new SLHA Block \Code{ChargedHiggsLHC13} containing
		various charged Higgs production cross sections (in pb, arbitarity
		values). The cross sections for charged Higgs production in association
		with one or two light flavor quarks ($u$, $d$, $s$) are generally
		combined to inclusive production processes containing one or two
		untagged jets. For the vector boson fusion process we set both quark PDG
		numbers to 1 in order to differentiate it from the other
		quark-associated production processes. All cross sections correspond to
		the sum of $H^+$ and $H^-$ production, hence, all PDG numbers are taken
		to be positive (except for $H^+H^-$ production).}
	\label{tab:SLHAcharged}
\end{table}

In \HBv{4} the \emph{squared} SM-normalized effective Higgs couplings to bosons
and third generation fermions were provided in the two SLHA blocks
\Code{HiggsBoundsCouplingInputBosons} and
\Code{HiggsBoundsCouplingInputFermions}, respectively. Since \HBv{5} requires
the sign information for the effective couplings, we have replaced these blocks
by very similar blocks named \Code{HiggsCouplingsBosons} and
\Code{HiggsCouplingsFermions} containing the \emph{non-squared}, sign sensitive
effective Higgs couplings as described in \cref{sec:theo}. In case only the old
blocks are specified in the SLHA input file for \HBv{5}, the effective Higgs
couplings are taken to be the \emph{positive} square-root of the given values.

In addition, we introduced an SLHA input block containing the hadronic cross
sections for direct charged Higgs boson production. In absence of a
corresponding SLHA convention, we call these input blocks
\Code{ChargedHiggsLHC8} and \Code{ChargedHiggsLHC13} for the predictions for the
LHC at \SIlist{8;13}{\TeV}, respectively.\footnote{Corresponding blocks for the
	Tevatron and the LHC at \SI{7}{\TeV} are irrelevant because no charged Higgs
	searches for these production processes have been performed.} The first three
columns specify the final state particle PDG numbers in increasing order (modulo
a sign in case of anti-particles). In case of a two-body final state the first
column is filled by a zero. The fourth column gives the cross section in~pb. An
example (employing the particle spectrum of a 2HDM Higgs sector) for one of
these SLHA blocks is given in \cref{tab:SLHAcharged}.

\section{Summary}
\label{sec:summary}
This paper documents a major update of the public \texttt{Fortran} code \HB
which tests general BSM models against exclusion limits from LEP, Tevatron and
LHC Higgs searches. We have presented the theoretical input framework of
\HBv{5}, which has been significantly extended to allow predictions for all
current and many potential future Higgs search channels. In particular this
extension adds sub-channels for several production modes --- such as $q\bar{q}$-
and $gg$-induced $Zh$ production --- that may be kinematically separable, and
incorporates flavor-violating decay modes and decays into BSM particles. The
charged Higgs boson input framework has also been extended by many different
direct production processes, some of which are already probed at the LHC.

We discussed the main experimental input for \HB\ --- the (nearly)
model-independent upper cross section limits --- and the possible limitations of
their application to BSM models. In fact, many of these limitations in current
search results can be overcome if more detailed information, in particular on
the signal composition in terms of Higgs production processes, is released
publicly by the experimental collaborations. Therefore we suggested guidelines
for the publication of experimental search results that we deem essential for a
proper reinterpretation in terms of BSM models. These recommendations are in
line with and partly extend those presented in Ref.~\cite{Abdallah:2020pec}.

In many BSM models precise calculations for ``exotic'' production cross sections
are in many cases not readily available. Therefore, for two of the important
production modes --- neutral Higgs production in association with a massive
gauge boson and top-associated charged Higgs production --- we have added
model-independent parameterizations of existing calculations. The effective
coupling (or scale factor) approximations of the $Zh$ and $W^\pm h$ production
cross sections are based on results obtained with the code
\vhatnnlo~\cite{Brein:2012ne,Harlander:2018yio} and include CP-sensitive
contributions. The $tH^+$ cross section parametrizes the precise calculations in
the 2HDM~~\cite{Berger:2003sm, Dittmaier:2009np, Flechl:2014wfa,
Degrande:2015vpa, deFlorian:2016spz, Degrande:2016hyf} through model-independent
coupling scale factors.

In most searches for additional Higgs bosons the final result provided by the
experimental collaborations is a --- potentially multidimensional --- upper
limit on the cross section at \CL{95} as a function of the relevant kinematic
variables, \ie the masses and widths of the involved particles. However, for the
neutral Higgs boson searches in the $\tau^+\tau^-$ final state a simplified
exclusion likelihood was provided by both the ATLAS and CMS collaborations.
These likelihoods are implemented in \HB, and the resulting likelihood value is
made available for use in model fits. We encourage the release of such
likelihoods also for other experimental search channels in the
future~\cite{Abdallah:2020pec}. We have improved the derivation of \CL{95}
limits from provided likelihood information by using the \CLs method and
presented a procedure to prevent overexclusion in case of excesses in the
searches. A validation of the likelihood implementation in the $M_h^{125}$
scenario of the MSSM using these new techniques found very good agreement with
the official results from CMS (ATLAS) taking into account (ignoring) the
model-dependent theoretical uncertainties on the signal rates.

\HBv{5} also involves substantial technical changes. The code is now available in a public git repository at
\begin{center}
	\url{https://gitlab.com/higgsbounds/higgsbounds}
\end{center}
with updates being released whenever new analyses have been implemented. We have
modernized the build system to use \texttt{CMake} and added a \texttt{C}
interface to make it easier for other codes to link to \HB. A technical
description of the user subroutines and further details on the code are given in
the online documentation at
\begin{center}
	\url{https://higgsbounds.gitlab.io/higgsbounds} .
\end{center}

\clearpage

\section*{Acknowledgments}

We thank Henning Bahl, Viviana Cavaliere, Andrew Gilbert, Artur Gottmann, Robert Harlander, Stefan
Liebler, Max Maerker, Bill Murray, Jana Schaarschmidt, Lukas Simon, Pietro Slavich, Roger
Wolf, Hanfei Ye and Lei Zhang for helpful
comments and discussions. We are grateful for earlier contributions from the former \HB team members Oliver Brein,
Oscar St{\aa}l, and Karina Williams.
The work of S.H.\ is supported in part by the
Spanish Agencia Estatal de Investigaci{\' o}n (AEI) and the EU Fondo Europeo
de Desarrollo Regional (FEDER) through the project FPA2016-78645-P,
in part by the MEINCOP Spain under contract FPA2016-78022-P,
in part by the “Spanish Red Consolider MultiDark” FPA2017-90566-REDC
and in part by
the AEI through the grant IFT Centro de Excelencia Severo Ochoa SEV-2016-0597.
T.S.\ and G.W.\ acknowledge support
by the Deutsche Forschungsgemeinschaft (DFG, German Research
Foundation) under Germany's Excellence Strategy -- EXC 2121 ``Quantum
Universe'' -- 390833306.
J.W.\ has been funded by the European Research Council (ERC) under the European
Union's Horizon 2020 research and innovation programme, grant agreement No
668679.

\begin{multicols}{2}[\printbibheading]
	\renewcommand*{\bibfont}{\small}
	\printbibliography[heading=none]
\end{multicols}

\end{document}